\RequirePackage{snapshot}
\documentclass[fleqn]{vch-book}
\usepackage{makeidx}
   \makeindex
\usepackage{graphicx}
\usepackage{epsf}
\usepackage{amsmath}
\usepackage{url}

\usepackage{latexsym}
\usepackage{amssymb}
\usepackage{bm}

\usepackage{booktabs}
\usepackage{parskip}

\usepackage{cite}
\usepackage[T1]{fontenc}
\usepackage[latin1]{inputenc}
\usepackage{times}
\usepackage{tabularx}
\usepackage{subfigure}
\usepackage{verbatim}

\usepackage{vch-adds}
\renewcommand{\ISBN}{3-527-40473-2} 
\renewcommand{\copyrightyear}[1]{2005}
\renewcommand{\copyrighttitle}[1]{%
  \textit{Phase Transitions in Combinatorial Optimization Problems.} Alexander K. Hartmann and Martin Weigt}

\showhyphens{propagation anisotropies parametrized dimensional}

\makeatletter
\@clubpenalty\clubpenalty
\makeatother

\begin{document}

\newcommand{\change}[1]{#1}
\newcommand{\changeM}[1]{#1}
\newcommand{\myvec}[1]{\underline{#1}}
\newcommand{\myvecB}[1]{{\bf #1}}
\newcommand{\mymatrix}[1]{\underline{\underline{#1}}}
\newcommand{\qed}{\hfill\ensuremath{\rm QED}}
\newcommand{\s}{s}
\newcommand{\myindex}[1]{\rule{5pt}{5pt}\index{#1}}

\newcommand{\narrowcaption}[2]{
\begin{minipage}{0.9\textwidth}
\vchcaption{#1}
\label{#2}
\end{minipage}}

\newlength{\boxlength}
\setlength{\boxlength}{\textwidth}
\addtolength{\boxlength}{-2\parindent}

\newenvironment{example}[1]
{\noindent\rule{\textwidth}{0.5mm}\vspace{-0.1cm}\begin{quote}
\underline{Example}: #1\\[0.1cm]}
{\hfill$\Box$\\
\end{quote}\vspace*{-0.7cm}\rule{\textwidth}{0.5mm}\\[0.0cm]}

\newenvironment{exampleNOL}[1]
{\rule{\textwidth}{0.5mm}\vspace{-0.1cm}\begin{quote}
\underline{Example}: #1\\[0.1cm]}
{\hfill$\Box$\\
\end{quote}}

\newenvironment{algorithm}[1]
{
\begin{tabbing} xx \= xx \= xx \= xx \= xx \= xx \= xx \=
  xxxxxxxxxxxxxxxxxxxxx  \= xxx \kill
 \textbf{algorithm} #1\\
 \textbf{begin}\\
}
{
 \textbf{end}\\
 \end{tabbing}
}

\newenvironment{procedure}[1]
{
\begin{tabbing} xx \= xx \= xx \= xx \= xx \= xx \= xx \=
  xxxxxxxxxxxxxxxxxxxxxx \= xxx  \kill
 \textbf{procedure} #1\\
 \textbf{begin}\\
}
{
 \textbf{end}\\
 \end{tabbing}
}

\newenvironment{definition}[1]
{\vspace{.5\baselineskip}\noindent
{\textbf{Definition:} \textit{#1}\\[4pt]}
}{\vspace{.5\baselineskip}}

\newenvironment{proof}
{\vspace{.5\baselineskip}\noindent\textbf{Proof:}\\[4pt]}
{\hfill QED\vspace{.5\baselineskip}}


\newcommand{\ii}[1]{{\it #1}}


\title{Phase Transitions in Combinatorial Optimization Problems}
\author{Alexander K. Hartmann and Martin Weigt}
\maketitle

\thispagestyle{empty}
\vspace*{5cm}
\begin{center}
\end{center}

\pagestyle{headings}

\setcounter{page}{24}
\setcounter{chapter}{2}
\chapter{Introduction to graphs}
\label{chap:graphs}
\thispagestyle{copyrightpage}
\index{graph|(}

The next \change{three} sections give a short introduction to graph theory and
graph algorithms. The first one deals with the basic definitions and concepts,
and introduces some graph problems. \change{The second one is dedicated to
  some fundamental graph algorithms. In the third one}, we will discuss random
graphs, which will be of fundamental importance throughout this course.

Let us begin by  mentioning some books related to graph theory. 
\index{textbooks} All of 
them go well beyond everything we will need concerning graphs:
\begin{itemize}
\item Gary Chartrand, \emph{Introductory graph theory}, Dover Publ.
  Inc., New York, 1985.\\
  This little paperback contains a nice, easy-to-read introduction to graph
  theory. Every chapter is based on ``real-world'' examples, which are mapped
  to graph problems. It is a good book for everyone who wishes to know more
  about graphs without working through a difficult mathematical book.
\item B\'ela Bollob\'as, \emph{Modern Graph Theory}, Springer, New York
  1998. \label{page:literature}\\
  Written by one of the leading graph theorists, this book covers the
  \change{first and third section} (and much more). It is very good and
  complete, but mathematically quite difficult.
\item Robert Sedgewick, \emph{Algorithms in C: Graph Algorithms}, 
  Addison Wesley, Boston 2002.\\
  This book collects many graph algorithms, including extensive
  descriptions and proofs of their correctness. 
\end{itemize}

\section{Basic concepts and graph problems}
\label{sec:3.1}

\subsection{The bridges of K\"onigsberg and Eulerian graphs}
\label{subsec:graph-defs}

The  earliest use of graph theoretical methods probably goes back to
the 18th century. At this time, there were seven bridges crossing the
river Pregel in the town of K\"onigsberg. The folks had long amused
themselves with the following problem: Is it possible to walk through
the town using every bridge just once, and returning home at the end?
The problem was solved by Leonhardt Euler 
\index{Euler} (1707--1783) in 1736 by
mapping it to a graph problem and solving it for arbitrary graphs
\cite{Eu}, i.\,e., for arbitrary towns, city maps, etc. In the case of
K\"onigsberg, he had to draw the slightly disappointing consequence that
no such round-walk existed.

Figure~\ref{fig:euler} shows the river Pregel, K\"onigsberg was situated
on both sides and both islands. The seven bridges are also shown. The
mapping to a graph is given below. Every river side, island and bridge
is assigned a \emph{vertex}, \index{vertex} \change{drawn as a circle,} 
two vertices are connected by an \emph{edge}, 
\change{drawn as a line,}
 if they are also physically connected. To give a more precise
description, we have to introduce the basic graph-theoretical
terminology which is summarized in the definitions below.

\begin{vchfigure}[htb]
\begin{center}
\scalebox{0.5}{\includegraphics{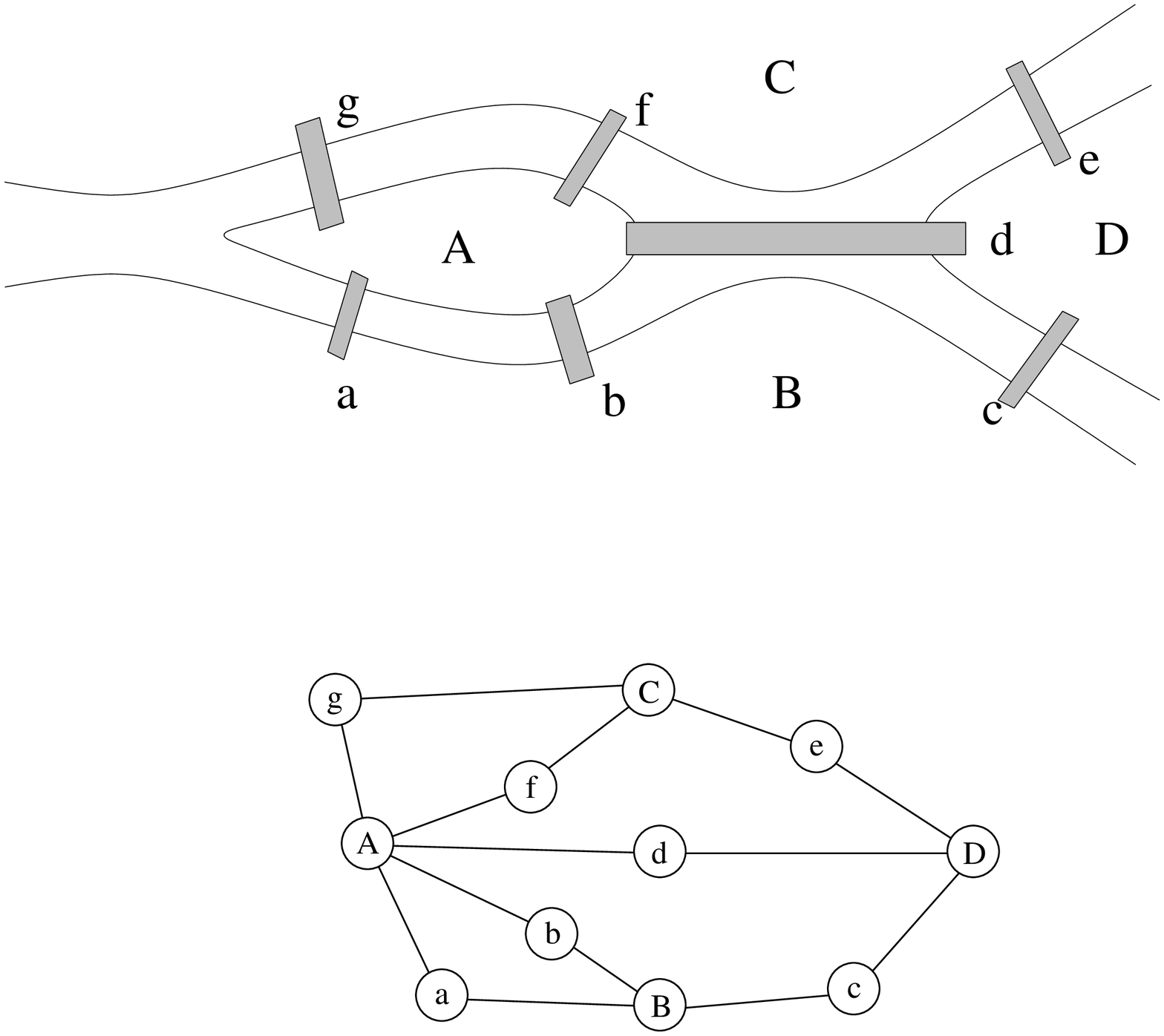}}
\end{center}
\vchcaption{The bridges of K\"onigsberg and its graph
  representation. \change{Vertices are denoted by circles, edges by lines.}}
\label{fig:euler}
\end{vchfigure}

\textbf{Basic definitions:}
\begin{itemize}
\item \change{An \emph{(undirected) graph} 
\index{graph|ii}\index{undirected graph|ii}\index{graph!undirected|ii} 
$G=(V,E)$ is given by its
 \emph{vertices} \index{vertex|ii} $i\in V$ and 
its undirected \emph{edges} \index{edge|ii} $\{i,j\}\in E
 \subset V^{(2)}$.  Note that both $\{i,j\}$ and $\{j,i\}$ denote the
 same edge.}
\item The \emph{order} \index{graph!order of}\index{order}
$N = |V|$ counts the number of vertices.
\item The \emph{size} \index{size of graph}\index{graph!size}
$M = |E|$ counts the number of edges.
\item Two vertices $i,j\in V$ are \emph{adjacent / neighboring} 
\index{adjacent} \index{neighbour} if
  $\{i,j\}\in E$.
\item The edge $\{i,j\}$ is \emph{incident} \index{incident edge}
\index{edge!incident} to its end vertices $i$ and 
  $j$.
\item The \emph{degree} \index{degree|ii} \index{vertex!degree}
${\rm deg}(i)$ of vertex $i$ equals the number
  of adjacent vertices. Vertices of zero degree are called \emph{isolated}.
\index{isolated vertex}\index{vertex!isolated}
\item A graph $G'=(V',E')$ is a \emph{subgraph} \index{subgraph} 
of $G$ if $V'\subset
  V,\ E'\subset E$.

\item A graph $G'=(V',E')$ is an \emph{induced subgraph} 
\index{induced subgraph}\index{subgraph!induced} of $G$ if
  $V'\subset V$ and  $E' = E\cap (V')^{(2)}$, i.\,e., $E'$ contains all
  edges from $E$ connecting two vertices in $V'$.

\item \change{A subgraph $G'=(V',E')$ is a \emph{path} \index{path}
of $G$ if it has
  the form $V'=\{i_0,i_1,\ldots,i_l\}$, \hbox{$E'=\{\{i_0,i_1\},\{i_1,i_2\},
  \ldots,\{i_{l-1},i_l\}\}$}. The \emph{length} 
\index{length of path}\index{path!length} of the path is $l=|E'|$.
  $i_0$ and $i_l$ are called \emph{end points}. \index{end point}
The path goes from
  $i_0$ to $i_l$ and vice versa. One says \emph{$i_0$ and $i_l$ are
  connected by the path}. \index{connected vertices}
\index{vertex!connected}
Note that, within a path, each vertex
  (possibly except for the end points) is ``visited'' only once.}

\item \change{ A path with $i_0=i_l$, i.\,e., a \emph{closed path}, 
\index{closed path}\index{path!closed} is called
    a \emph{cycle}. \index{cycle}}

\item \change{A sequence of edges $\{i_0,i_1\}$, $\{i_1,i_2\},
  \ldots, \{i_{l-1},i_l\}$ is called a \emph{walk}. \index{walk}
Within a
  walk some vertices or edges may occur several times.}

\item \change{A walk with pairwise distinct edges is called a \emph{trail}. 
\index{trail} Hence a trail is also a subgraph of~$G$.}

\item \change{A \emph{circuit} \index{circuit} 
is trail with coinciding end points,
  i.\,e., a \emph{closed} trail. \index{closed trail}\index{trail!closed} 
(NB: cycles are circuits, but not vice
  versa, because a circuit may pass through several times through the
  same vertex).}

\item The graph $G$ is \emph{connected} \index{connected graph}
\index{graph!connected} if all pairs $i,j$ of vertices 
  are connected by paths.

\item The graph $G'=(V',E')$ is a \emph{connected component} 
\index{connected component|ii}\index{component!connected|ii} of $G$ if
  it is a connected, induced subgraph of $G$, and there are no edges
  in $E$ connecting vertices of $V'$ with those in $V\setminus
  V'$.

\item \change{The \emph{complement graph} \index{complement graph}
\index{graph!complement} $G^C=(V,E^C)$ has edge set
  $E^C = V^{(2)}\setminus E = \{\{i,j\}\mid\{i,j\}\notin E\}$. It is
  thus obtained from $G$ by
  \label{page:graph_defs} connecting all vertex pairs by an edge, which 
  are not adjacent in $G$ and disconnecting all vertex pairs, which
  are adjacent in $G$.}

\item A \emph{weighted graph} \index{weighted graph|ii}
\index{graph!weighted|ii} $G=(V,E,\omega)$ is a graph with edge
  weights $\omega: E \to \mathbb R$.
\end{itemize}

\change{
\begin{example}{Graphs}
We consider the graph shown in Fig.~\ref{fig:euler}. It can be
written as $G=(V,E)$ with
\begin{eqnarray*}
V & = & \{A,B,C,D,a,b,c,d,e,f,g\}\\
E & = & \{\{A,a\}, \{A,f\}, \{A,d\}, \{A,f\}, \{A,g\}, 
\{B,a\}, \{B,b\}, \{B,c\}, \\
& & \,\,\,\{C,e\}, \{C,f\}, \{C,g\}, 
\{D,c\}, \{D,d\}, \{D,e\}, \}\,.
\end{eqnarray*}
Hence, the graphs has $|V|=11$ vertices and $|E|=14$ edges. Since
$\{D,e\}\in E$, the vertices $D$ and $d$ are adjacent. Vertex $d$ has
degree deg$(d)=2$, while vertex $A$ has degree 5. 

For example, $G'=(V',E')$ with $V'=\{A,g,C,e,D\}$ and 
$E'=\{\{A,g\}$, $\{g,C\}$, $\{C,e\}$, $\{e,D\}, \}$ is a path from $A$ to $D$.
$G$ is connected,
because all vertices are connected by paths to all other vertices.
The sequence of edges $\{B,c\},\{c,D\},\{D,c\}$ is a walk, but it
does not correspond to a path, because some vertices are visited
twice. The sequence of edges $\{A,b\}$, $\{b,B\}$, $\{B,a\}$, $\{a,A\}$, 
$\{A,g\}$, $\{g,C\}$, $\{C,f\}$, $\{f,A\}$ is a trail, 
in particular it is a circuit.
\end{example}
}

Going back to the problem of the people from K\"onigsberg, formulated in
graph-theoretical language, they were confronted with the following
question:

\noindent
\textbf{EULERIAN CIRCUIT:} \ Given 
\index{EULERIAN CIRCUIT} \index{circuit!Eulerian}
a graph, is there a circuit using 
every \emph{edge} exactly once?

The amazing point about Euler's proof is the following: The existence
of a Eulerian circuit -- which obviously is a global object -- can be
decided looking to purely local properties, in this case to the
vertex degrees.

\noindent
\textbf{Theorem:} \emph{A connected graph $G=(V,E)$ is Eulerian (has an 
  Eulerian cycle) iff all vertex degrees are even.} \index{degree}

\begin{proof}
$(\rightarrow)$ This direction is trivial. Following the Eulerian
circuit, every vertex which is entered is also left, every time using
previously unvisited edges. All degrees are consequently even.

$(\leftarrow)$ The other direction is a bit harder. We will proof it
by induction on the graph size $M$, for arbitrary graphs having only
even vertex degrees.

The theorem is obviously true for $M=0$ (one isolated vertex) and
$M=3$ (triangle) which are the simplest graphs with even degrees.

Now we take any $M>0$, and we assume the theorem to be true for all 
graphs of size smaller than $M$. We will show that the theorem is
also true for a connected graph $G$ of size $M$ having only even
degrees.

Because of $M>0$, the graph $G$ is non-trivial, because of the even
degrees it contains vertices of degree at least 2. Then $G$ contains a
cycle, which can be seen in the following way: Start in any vertex and
walk along edges. Whenever you enter a new vertex, there is at least
one unused edge which you can use to exit the vertex again. At a
certain moment you enter a vertex already seen (at most after $M$
steps), the part of the walk starting there is a cycle.

Every cycle is also a circuit. Consider now a circuit $C=(G',E')$ of
maximal size $|E'|$. If $C$ is a Eulerian circuit in $G$, everything
is OK. If not, we have $|E'| < |E|$, and the subgraph $H=(V,E\setminus
E')$ has at least one non-trivial connected component $H'$. A circuit 
has even degrees, thus $H$ and all its connected components have even
degrees. Since $H'$ has size $<M$, it has an Eulerian circuit
which can be added to $C$, violating the maximality of $C$. Thus, $C$
has to be an Eulerian circuit, and the proof is complete.
\end{proof}

Going back to K\"onigsberg, we see that there are vertices of odd degrees. No
Eulerian circuit can exist, and the inhabitants have either to skip bridges or
to cross some twice if they walk through their town.

\change{In the above definitions for undirected graphs, 
edges have no orientation. This is
different for \emph{directed} graphs (also \index{digraph}
called \emph{digraphs}). 
Most definitions for directed
graphs are the same as for undirected graphs. Here we list some
differences and additional definitions.
\begin{itemize}
\item A \emph{directed graph} \index{graph!directed|ii}
\index{directed graph|ii}
$G=(V,E)$ is similar to an undirected
  graph, except that the edges (also called \emph{arcs} 
\index{arc} in this case)
  $(i,j)\subset V\times V$ are ordered pairs of vertices.

\item The \emph{outdegree} \index{outdegree}\index{degree!out--}
 $d_{\rm out}(i)$ is the number of outgoing edges
    $(i,j)$.

\item The \emph{indegree} \index{indegree}\index{degree!in--}
 $d_{\rm in}(i)$ is the number of incoming edges
    $(j,i)$.
\item A directed \emph{path} \index{directed path}\index{path!directed}
is defined similarly
 to a path for undirected graphs, except that
  all edges must have the same ``forward'' 
orientation along the path. Formally, a
  path form $i_0$ to $i_l$ is a subgraph $G'=(V',E')$ with 
$V'=\{i_0$,$i_1,\ldots,i_l\}$, $E'=\{(i_0,i_1)$,$(i_1,i_2)$,
\ldots,$(i_{l-1},i_l)\}$.

\item A directed graph $G=(V,E)$ is called \emph{strongly connected}
\index{strongly connected graph}\index{graph!strongly connected}%
  if for all pairs of vertices $i,j$, there exists a directed path in
  $G$ from $i$ to $j$ and a path from $j$ to $i$. A \emph{strongly
  connected component} \index{strongly connected component|ii}
\index{component!strongly connected|ii} 
\index{SCC|see{strongly connected component}}
(SCC) of a directed graph is a strongly connected
  subgraph of maximal size, i.\,e., it cannot be extended by adding
  vertices/edges while still being strongly connected.
\end{itemize}
}

\subsection{Hamiltonian graphs}
\index{Hamiltonian graph|(}\index{graph!Hamiltonian|(}
\index{Hamiltonian-cycle problem|(}\index{problem!Hamiltonian-cycle|(}%

Imagine you have to organize a diplomatic dinner. The main problem is placing
the $N$ ambassadors around a round table: There are some countries which are
strong enemies, and during the dinner there is a risk that the diplomats will
not behave very diplomatically. You therefore prefer to place only those
ambassadors in neighboring seats who belong to countries which have peaceful
relations.

\begin{vchfigure}[htb]
\begin{center}
\scalebox{0.5}{\includegraphics{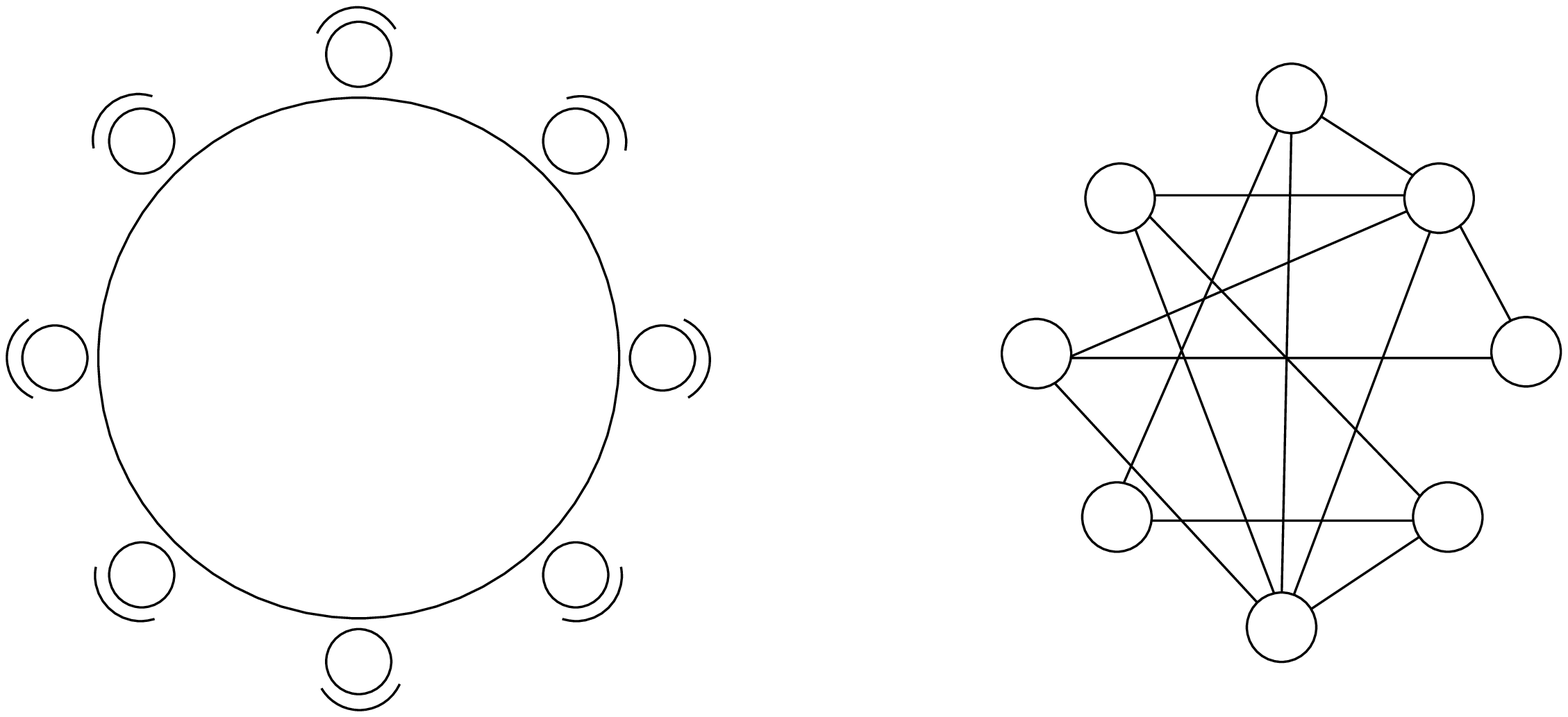}}
\end{center}
\vchcaption{Left: The round table where the diplomats have to be
  placed.
  Right: Graph of relations, only ambassadors with positive
  relations are connected and can be neighbors at the table.}
\label{fig:diplomats}
\end{vchfigure}

Again, this problem can be mapped to a graph-theoretical question.
The corresponding graph is the ``graph of good relations''. The
vertices are identified with the ambassadors, and any two having good
relations, i.\,e., potential neighbors, are connected by an edge,
see Fig.~\ref{fig:diplomats}. The problem now reduces to finding a
cycle of length $N$:

\noindent
\textbf{HAMILTONIAN CYCLE:} (HC) 
\index{Hamiltonian-cycle problem}\index{problem!Hamiltonian-cycle}%
\index{HC|see{Hamiltonian-cycle problem}}
Is there a cycle through a given graph 
such that every \emph{vertex} is visited exactly once?

At first, this problem seems to be similar to the search
for an Eulerian circuit, but it is not. It belongs to the NP-complete
problems (see Chap.~4) and is, in general, not
decidable from local considerations. An exception is given in the case
of highly connected  graphs:

\noindent
\textbf{Theorem:} \emph{If $G=(V,E)$ is a graph of order $N\geq 3$ such
  that ${\rm deg}(i)\geq N/2$ for all vertices in $V$, then $G$ is
  Hamiltonian (i.\,e., it has a Hamiltonian cycle).}

We do not give a proof, we just state that the theorem is not
sufficient to say anything about our specific sample given in 
Fig.~\ref{fig:diplomats}. There are vertices of degree 2 and 3,
which are smaller than $N/2=4$.

The search for Hamiltonian cycles is closely related to the already
mentioned traveling-salesman problem (TSP), 
\index{traveling-salesman problem}\index{problem!traveling-salesman} 
which is defined on a
\emph{weighted} graph. \index{weighted graph}\index{graph!weighted}
The \emph{total weight of a subgraph} is the sum
of the weights of all its edges. In TSP we are looking for the
Hamiltonian cycle of minimum weight.
\index{Hamiltonian graph|)}\index{graph!Hamiltonian|)}
\index{Hamiltonian-cycle problem|)}\index{problem!Hamiltonian-cycle|)}%

\subsection{Minimum spanning trees}
\label{subsec:spanning}

Let us come to the next problem. This time you have to plan a railway
network in a country having no trains but many towns which are to be
connected. The country has financial problems and asks you to design
the cheapest possible network, providing you with the information of which
pairs of towns can be directly connected, and at what cost -- i.\,e.
they give you a weighted graph. You must find a connected
subgraph containing each town to enable travel from one town to
another. It is also clear that loops increase the total cost. The
most expensive edge in a loop can be deleted without disconnecting
some towns from others. But how can one find the globally optimal
solution?

Before answering this, we have to go through some definitions:

\begin{definition}{tree, forest} 
A \emph{tree} \index{tree} is a connected cycle-free graph.\\ 
A \emph{forest} \index{forest} is a graph which has only trees as connected
components, see, e.\,g., Fig.~\ref{fig:forest}.
\end{definition}

\begin{vchfigure}[htb]
\begin{center}
\scalebox{0.5}{\includegraphics{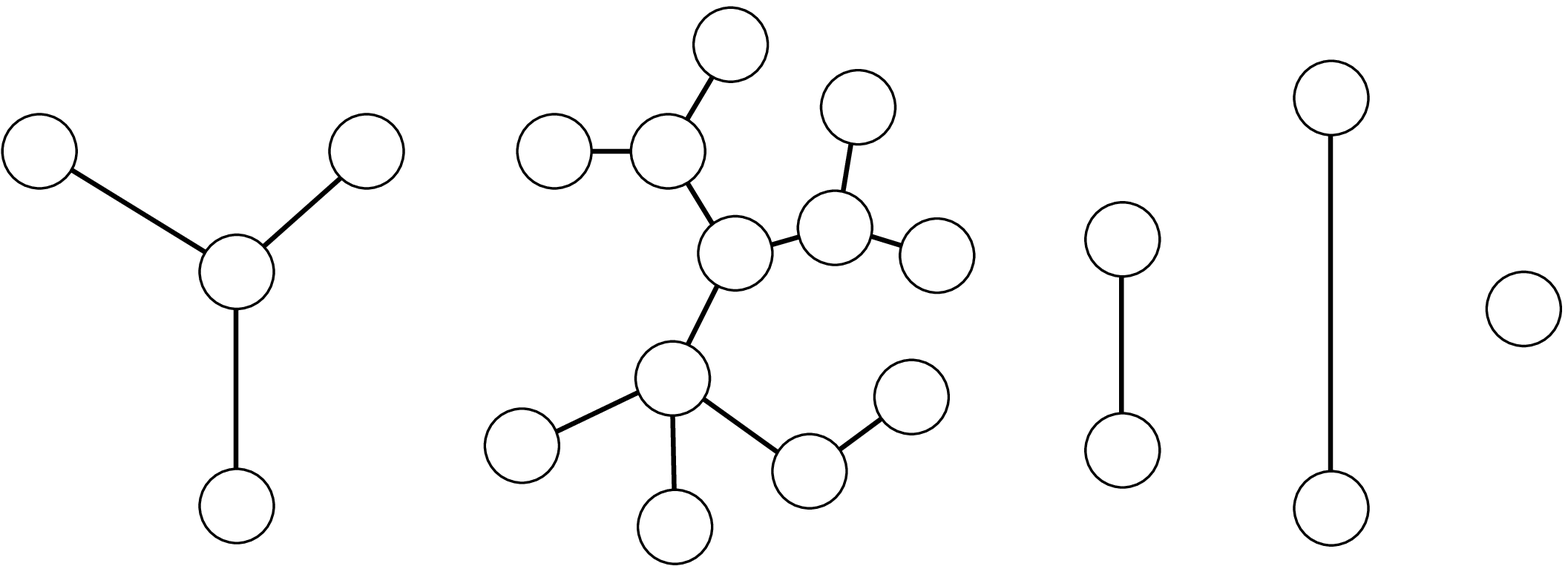}}
\end{center}
\vchcaption{A forest is made of trees.}
\label{fig:forest}
\end{vchfigure}

Some easy-to-prove properties of trees are given here:

\noindent
\textbf{Theorem:} \emph{A tree of order $N$ has  size $M=N-1$.}

\begin{proof}
Start with one isolated vertex, add edges. Since there
are no cycles, every edge adds a new vertex. 
\end{proof}

\noindent
\textbf{Corollary:} \emph{Every tree of order $N\geq 2$ contains at
 least two vertices of degree 1 (leaves). \index{degree}} 
\index{leaf|ii}\index{graph!leaf|ii}

\begin{proof}
 Assume the opposite, i.\,e., that there are at least $N-1$
 vertices of degree at least 2. We thus find
\begin{equation*}\sum_{i\in V} {\rm deg}(i) \geq 2N-1 \end{equation*}
On the other hand, we have
\begin{equation*} \sum_{i\in V} {\rm deg}(i) = 2M \end{equation*}
since every edge is counted twice. This produces a contradiction to
the theorem. 
\end{proof}

To approach our railway-network problem, we need two further definitions.

\begin{definition} {(Minimum) spanning tree/forest}
\begin{itemize} 
\item Let $G=(V,E)$ be a connected graph of order $N$. A \emph{spanning
  tree} \index{spanning tree}\index{tree!spanning}
is a cycle-free subgraph of order $N$ having maximum size $N-1$,
i.\,e., the spanning tree is still connected.

\item If $G=(V,E,\omega)$ is weighted, a spanning tree of minimum weight is
called an \emph{minimum spanning tree} \index{minimum spanning tree} 
\index{spanning tree!minimum} (or \emph{economic spanning tree}).
\index{economic spanning tree}\index{spanning tree!economic}

\item For a general graph, a \emph{(minimum) spanning forest} 
\index{spanning forest}\index{forest!spanning}%
 \index{minimum spanning forest}\index{spanning forest!minimum}
is the
  union of all (minimum) spanning trees for its different connected components.
\end{itemize}
\end{definition}

Due to the maximum size condition of a spanning tree, 
the connected components \index{connected component}
\index{component!connected} of a graph and of its spanning
forest are the same.
Spanning trees can be constructed using algorithms which calculate
connected components, see Sec.~\ref{subsec:depth-breadth}.
Minimum spanning trees obviously fulfil all requirements to our
railway network, and they can be constructed in a very simple way
using Prim's algorithm, which is presented in Sec.~\ref{subsec:prims}.

\subsection{Edge covers and vertex covers}
\label{sec:vc-def}

\change{ Now we introduce edge and vertex covers. Both problems are
similar to each other, but they are fundamentally different with
respect to how easily they can
be solved algorithmically. This we have already seen for Eulerian circuits
and Hamiltonian cycles. The vertex-cover problem will serve as the
prototype example in this book. All concepts, methods and analytical
techniques will be explained using the vertex-cover problem.  We begin
here with the basic definitions, but additional definitions, related to
vertex-cover algorithms, are given in Sec.~6.1.

For a graph $G=(V,E)$, an \emph{edge cover}\/ \index{cover!edge}
\index{edge cover} is a subset $E'\subset E$ of edges such that each
vertex is contained in at least one edge $e\in E'$. Each graph which
has no isolated vertices has an edge cover, since in that case $E$
itself is an edge cover. A \emph{minimum edge cover}\/ \index{minimum
edge cover}\index{edge cover!minimum} is an edge cover of minimum
cardinality $|E^{'}|$. In Fig.~\ref{fig-cover} a graph and a minimum
edge cover are shown. A fast algorithm which constructs a minimum edge
cover can be found in Ref.~\cite{GRAPH-lawler76a}.

\begin{vchfigure}[ht]
\begin{center}
\scalebox{0.5}{\includegraphics{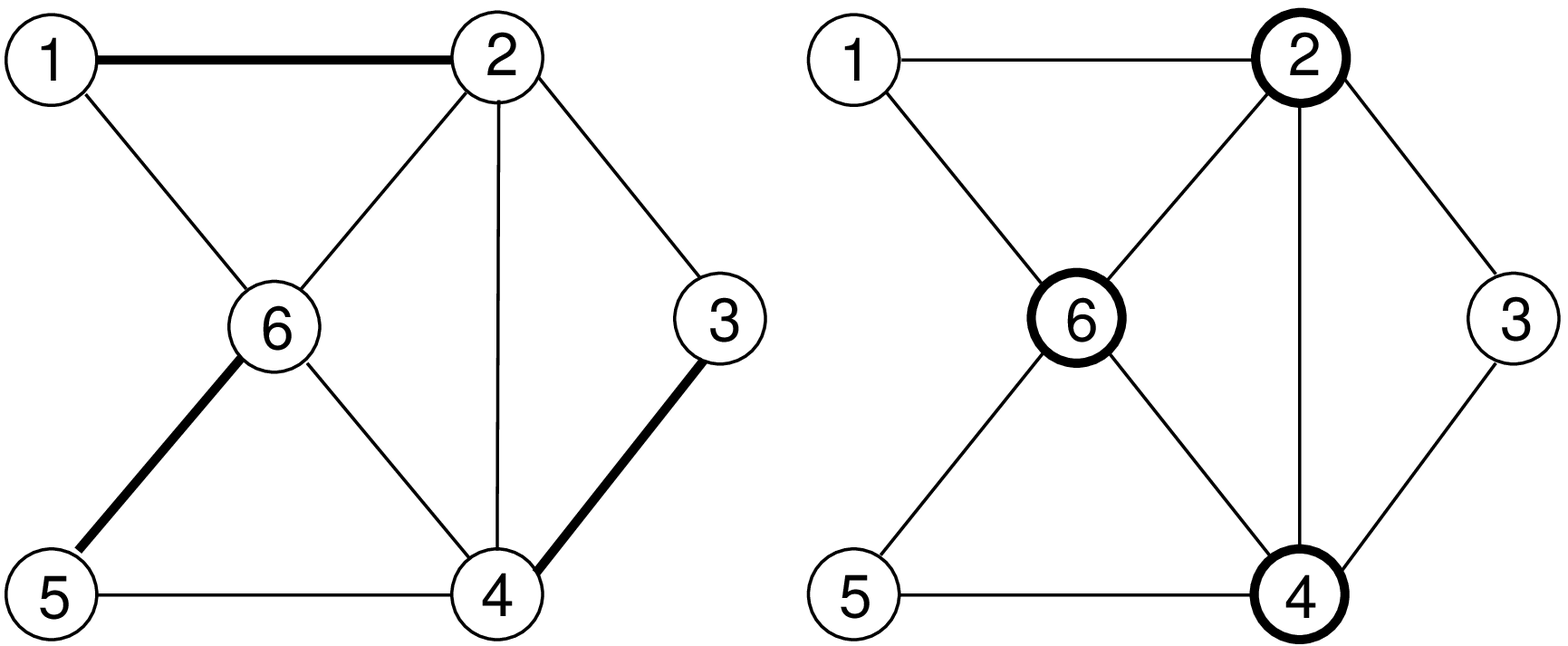}}
\vchcaption{A graph and a minimum edge cover (left) and a minimum vertex
  cover (right). The edges/vertices belonging to the cover are shown
  in bold.}
\label{fig-cover}
\end{center}
\end{vchfigure}

The definition of a \emph{vertex cover}\index{cover!vertex}
\index{vertex cover} is very similar. It is a subset $V'\subset V$ of
vertices such that each edge $e=\{i,j\}\in E$ contains at least one
vertex out of $V'$, i.\,e., $i\in V'$ or $j\in V'$. Note that $V$ itself
is always a vertex cover. A \emph{minimum vertex cover}\/ \index{minimum
vertex cover}\index{vertex cover!minimum} is an vertex cover of minimum
cardinality $|V^{'}|$.

Vertex covers are closely related to 
\emph{independent sets} \index{independent set} and \emph{cliques}. 
\index{clique|ii} An independent set of a graph $G=(V,E)$ is a subset
$I\subset V$ of vertices, such that for all elements $i,j\in I$, there
is no edge $\{i,j\}\in E$. A clique is a subset $Q\subset V$ of
vertices, such that for all elements $i,j\in Q$ there is an edge
$\{i,j\}\in E$.

\textbf{Theorem:} \emph{For a given graph $G=(V,E)$ and a 
subset $V'\subset V$ the following three
statements are equivalent.
\begin{itemize}
\item[(A)] $V'$ is a vertex cover of $G$.
\item[(B)] $V\setminus V'$ is an independent set of $G$.
\item[(C)] $V\setminus V'$ is a clique of the complement graph $G^C$
  (see definition on page \pageref{page:graph_defs}).
\end{itemize}
}

\begin{proof}
(A $\rightarrow$ B)\\ 
Let $V'$ be a vertex cover, and $i,j\in
 V\setminus V'$. We assume that there is an edge $\{i,j\}\in E$. Since
 $i,j\not\in V'$, this is an edge with both vertices not in $V'$, and
 $V'$ is not a vertex cover. This is a contradiction! Hence, there
 cannot be an edge $\{i,j\}\in E$, and $V\setminus V'$ is an
 independent set.

(B $\rightarrow$ C)\\ 
Let $V\setminus V'$ be an independent set, and
$i,j\in V\setminus V'$. By definition, there is no edge $\{i,j\}\in
E$, and so there is an edge $\{i,j\}\in E^C$. Therefore, $V\setminus
V'$ is a clique of $G^C$.

(C $\rightarrow$ A)\\ 
Let $V\setminus V'$ be a clique of $G^C$, and
$\{i,j\} \in E$. This means $\{i,j\}\not\in E^C$ by definition of
$G^C$. Thus, we have $i\not\in V\setminus V'$ or $j\not\in V\setminus
V'$ because $V\setminus V'$ is a clique. Hence, $i\in V'$ or
$j\in V'$. By definition of vertex cover, $V'$ is a vertex cover.
\end{proof}

The \emph{minimum vertex cover} is a vertex cover of minimum
cardinality.  From the theorem above, for a minimum vertex cover $V'$,
$V\setminus V'$ is a maximum independent set of $G$ and a maximum
clique of $G^C$.  In Fig.~\ref{fig-cover}, a graph together with its
minimum vertex cover is displayed. Related to the minimization problem
is the following decision problem, for given integer $K$:\\

\textbf{VERTEX COVER (VC):}  
\index{vertex-cover problem|ii}\index{problem!vertex-cover|ii} 
\index{VC|see{vertex-cover problem}} Does
a given graph $G$ have a vertex cover $V'$ with $|V^{'}|\le K$?

In Chap.~4 we will see that VC is a so-called NP-complete
\index{NP-complete}
problem, which means in particular that no fast algorithm for solving this
problem is known -- and not even expected to be able to be constructed. This
proposition and its relation to statistical physics will be the main subject
of this book.

}

\subsection{The graph coloring problem}
\label{subsec:coloring}
\index{graph!coloring|(}\index{coloring!graph|(}

The last problem which we discuss in this section is a scheduling problem.
Imagine you have to organize a conference with many sessions which, in
principle, can take place in parallel. Some people, however, want to
participate in more than one session. We are looking for a perfect
schedule in the sense that every one can attend all the sessions he
wants to, but the total conference time is minimum.

The mapping to an \change{undirected} graph is the following: The sessions are
considered as vertices. Two of them are connected by an edge whenever there is
a person who wants to attend both. We now try to \emph{color} the vertices in
such a way that no adjacent vertices carry the same color. Therefore we can
organize all those sessions in parallel, which have the same color, since
there is nobody who wants to attend both corresponding sessions. The optimal
schedule requires the minimum possible number of colors.

The problem is an analog to the coloring of geographic maps, where
countries correspond to vertices and common borders to edges. Again we do
not want to color neighboring countries using the same color.  The
main difference is the structure of the graph. The ``geographic'' one
is two-dimensional, without crossing edges -- the so-called \emph{planarity} 
\index{planar graph}\index{graph!planar} of the graph 
makes the problem easy. The first graph is
more general, and therefore more complicated. Again, the problem
becomes NP-complete. \index{NP-complete}

Let us be more precise:

\begin{definition}{$q$-coloring, chromatic number, $q$-core}
\begin{itemize}
\item A \emph{$q$-coloring} \index{q-coloring@$q$-coloring|ii}
of $G=(V,E)$ is a mapping $c: V \to
  \{1,\ldots,q\} $ such that $c(i)\neq c(j)$ for all edges $\{i,j\}\in 
  E$.

\item The minimum number of needed colors is called the 
\emph{chromatic number} \index{chromatic number} of $G$.

\item The maximal subgraph, where the minimum degree over all vertices
  is at least $q$, is called the
  \emph{$q$-core}. \index{q-core@$q$-core|ii}
\index{core|see{$q$-core}}
\end{itemize}
\end{definition}

The smallest graph which is not $q$-colorable is the 
\emph{complete graph} \index{complete graph}\index{graph!complete}
 $K_{q+1}$ of order $q+1$ and size $q(q+1)/2$, i.\,e., all pairs
of vertices are connected. Complete graphs are also called 
\emph{cliques}. \index{clique}
For $K_3$ this is a triangle, for $K_4$ a tetrahedron.

Before analyzing the $q$-coloring problem, a linear-time algorithm for
constructing the $q$-core is introduced. The idea of the algorithm is
easy, it removes all vertices of degree smaller than $q$, together
with their incident edges. The reduced graph may have new vertices of
degree smaller than $q$, which are removed recursively, until a subgraph of
minimum degree at least $q$ is obtained. For obvious reasons the
$q$-core is obtained: No vertex out of the $q$-core can be removed by
the algorithm, so the final state is at least as large as the
$q$-core. By the maximality of the latter, both have to be equal.  The
input to the algorithm presented below is the initial graph $G=(V,E)$,
the output is its $q$-core:

\index{q-core algorithm@$q$-core algorithm}
\index{algorithm!q-core@$q$-core}
\begin{algorithm}{core($V,E,q$)}
  \> \textbf{if} min$\{ {\rm deg}(i)| i\in V \} \geq q$ \textbf{then}\\
  \>\> \textbf{return} $G=(V,E)$;\\
  \> \textbf{else do}\\
  \>\> \textbf{begin}\\
  \>\>\> $U := \{i\in V | {\rm deg}(i)<q  \} $;\\
  \>\>\> $V := V\setminus U$;\\
  \>\>\> $E := E \cap V^{(2)}$;\\
  \>\>\> \textbf{return} core($V,E,q$);\\
  \>\> \textbf{end}\\
\end{algorithm}

\textbf{Theorem:} \emph{A graph is $q$-colorable if and only if its
$q$-core is $q$-colorable.}
\label{page:coloring}\index{q-core@$q$-core}\index{q-coloring@$q$-coloring}

\begin{proof}
($\rightarrow$) A coloring of the full graph is obviously a
coloring of every subgraph.

($\leftarrow$) Construct the $q$-core by the core-algorithm, but keep
track of the order in which vertices are removed. Now take any
$q$-coloring of the core. Reconstruct the full graph by adding the
removed vertices in inverse order. Directly after adding, the vertices
have degree smaller than $q$. Their neighbors use less than $q$
colors, i.\,e., the added vertex can be colored. In this way, every
coloring of the $q$-core can be recursively extended to a coloring of
the full graph. 
\end{proof}

\begin{vchfigure}[htb]
\begin{center}
\scalebox{0.5}{\includegraphics{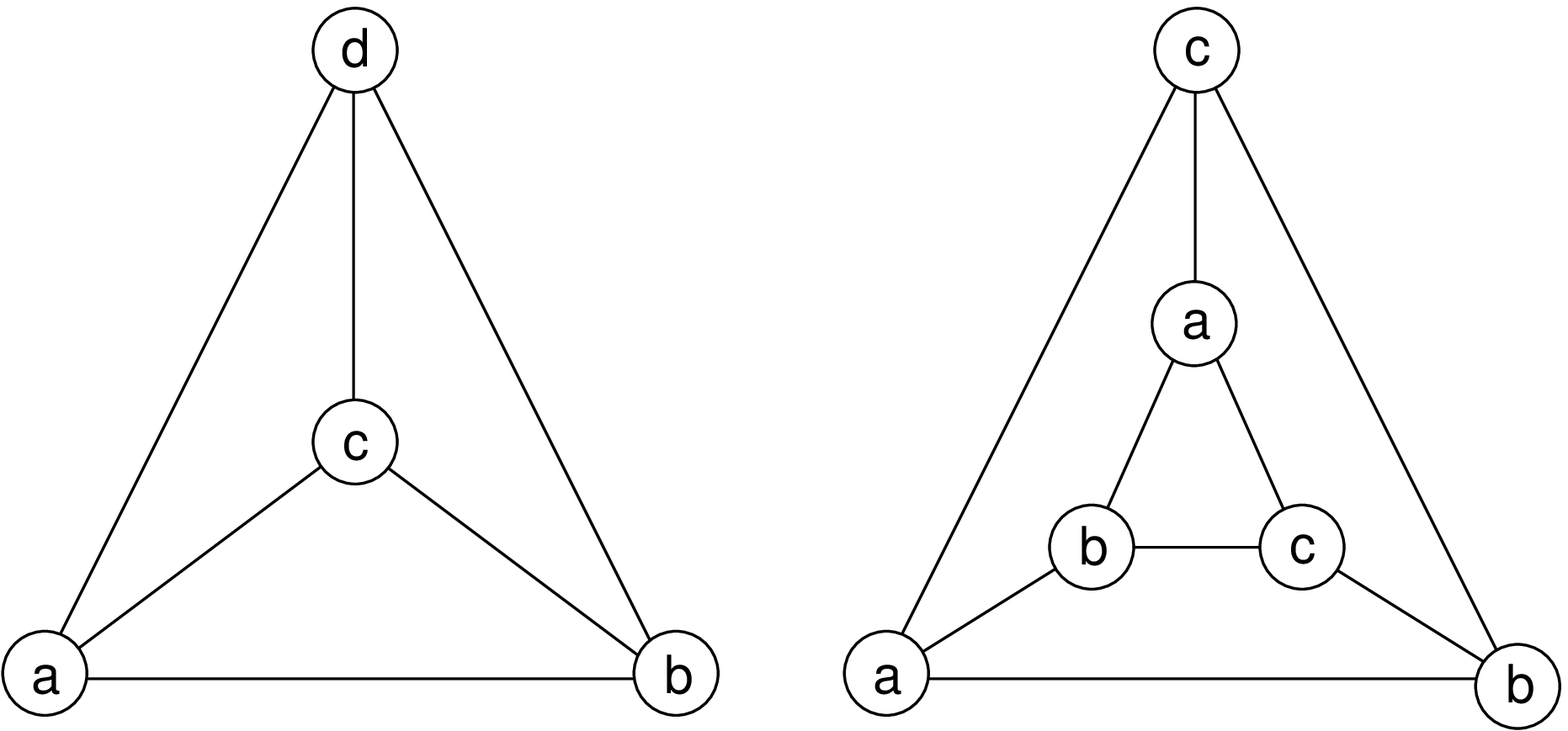}}
\end{center}
\vchcaption{The left graph is the tetrahedron $K_4$ which is not
  3-colorable, the letters indicate the different colors of a
  4-coloring. The right graph also has fixed degree 3, but is
  colorable with three colors.}
\label{fig:col}
\end{vchfigure}

This shows that the hard part of $q$-coloring a graph consists in
coloring its $q$-core, i.\,e., the existence of a non-trivial $q$-core is
necessary, but not sufficient for $q$-uncolorability, see, e.\,g., the
examples given in Fig.~\ref{fig:col}. This also leads to

\textbf{Corollary:}  \emph{Every graph having at most $q$ vertices of degree
at least $q$ is $q$-colorable.}
\index{q-core@$q$-core}\index{q-coloring@$q$-coloring}\index{degree}

In the beginning of this section we have also considered the coloring
of geographic maps -- or \emph{planar graphs}, i.\,e., graphs which can be
drawn in a plane without crossing edges. For these graphs,
the chromatic number can be found easily, using the following famous
theorem:

\textbf{Theorem:} \emph{Every planar graph is 4-colorable.}
\index{planar graph}\index{graph!planar}

This had already been conjectured in 1852, but the final solution was only
given in 1976 by Appel and Haken \cite{ApHa}. Their proof is, however,
quite unusual: Appel and Haken ``reduced'' the original question to
more than 2000 cases according to the graph structure. These cases
were finally colored numerically, using about 1200 hours of
computational time. Many mathematicians were quite dissatisfied by
this proof, but up to now nobody has come up with a ``hand-written'' one.

On the other hand, it is easy to see that 3 colors are in general not
sufficient to color a planar graph. The simplest example is the
4-clique $K_4$, which is planar, as can be seen in Fig.~\ref{fig:col}.
\index{graph!coloring|)}\index{coloring!graph|)}

\subsection{Matchings}

\change{ Given a graph $G=(V,E)$, \index{graph} a matching $M \subset
E$ is a subset of edges, such that no two edges in $M$ are incident to
the same vertex \cite{GRAPH-GGL,GRAPH-MatchTh}, i.\,e., for all vertices
$i\in V$ we have $i\in e$ for at most one edge $e\in M$.  An edge
contained in a given matching is called
\emph{matched},\index{edge!matched} other edges are
\emph{free}.\index{edge!free} A vertex belonging to an edge $e \in M$
is \emph{covered}\index{vertex!covered}\index{vertex!matched} (or
\emph{matched}) others are \emph{M-exposed}\/ \index{vertex!M-exposed}
\index{M-exposed vertex} (or \emph{exposed}\index{vertex!exposed} or
\emph{free}).\index{vertex!free}\index{free vertex|ii}If $e=\{i,j\}$
is matched, then $i$ and $j$ are called \emph{mates}.

A matching $M$ of maximum cardinality is called\/
\emph{maximum-cardinality matching}\/.\index{maximum-cardinality matching}
\index{matching!maximum-cardinality} A matching is 
\emph{perfect}\/\index{matching!perfect}\index{perfect matching}
if it leaves no exposed vertices, hence it is automatically a
maximum-cardinality matching. On the other hand, not every
maximum-cardinality matching is perfect.\index{mates}

The vertex set of a \emph{bipartite graph}\index{bipartite
  graph|ii}\index{graph!bipartite|ii} can be subdivided into two disjoint
sets of vertices, say $U$ and $W$, such that edges in the graph $\{i,j\}$ only
connect vertices in $U$ to vertices in $W$, \emph{with no edges}\/ internal to
$U$ or $W$. Note that nearest-neighbor hypercubic lattices are bipartite,
while the triangular and face-centered-cubic lattices are \emph{not}\/
bipartite.  Owing to the fact that all cycles on bipartite graphs have an even
number of edges, matching on bipartite graphs is considerably easier than
matching on general graphs. In addition, maximum-cardinality matching on
bipartite graphs can be easily related to the maximum-flow 
 \index{maximum flow}\index{flow!maximum} problem
\cite{GRAPH-rieger1998,GRAPH-alava2000,GRAPH-opt-phys2001}, see
Sec.~11.5.

\begin{example}{Matching}

\begin{vchfigure}[htb]
\begin{center}
\scalebox{0.5}{\includegraphics{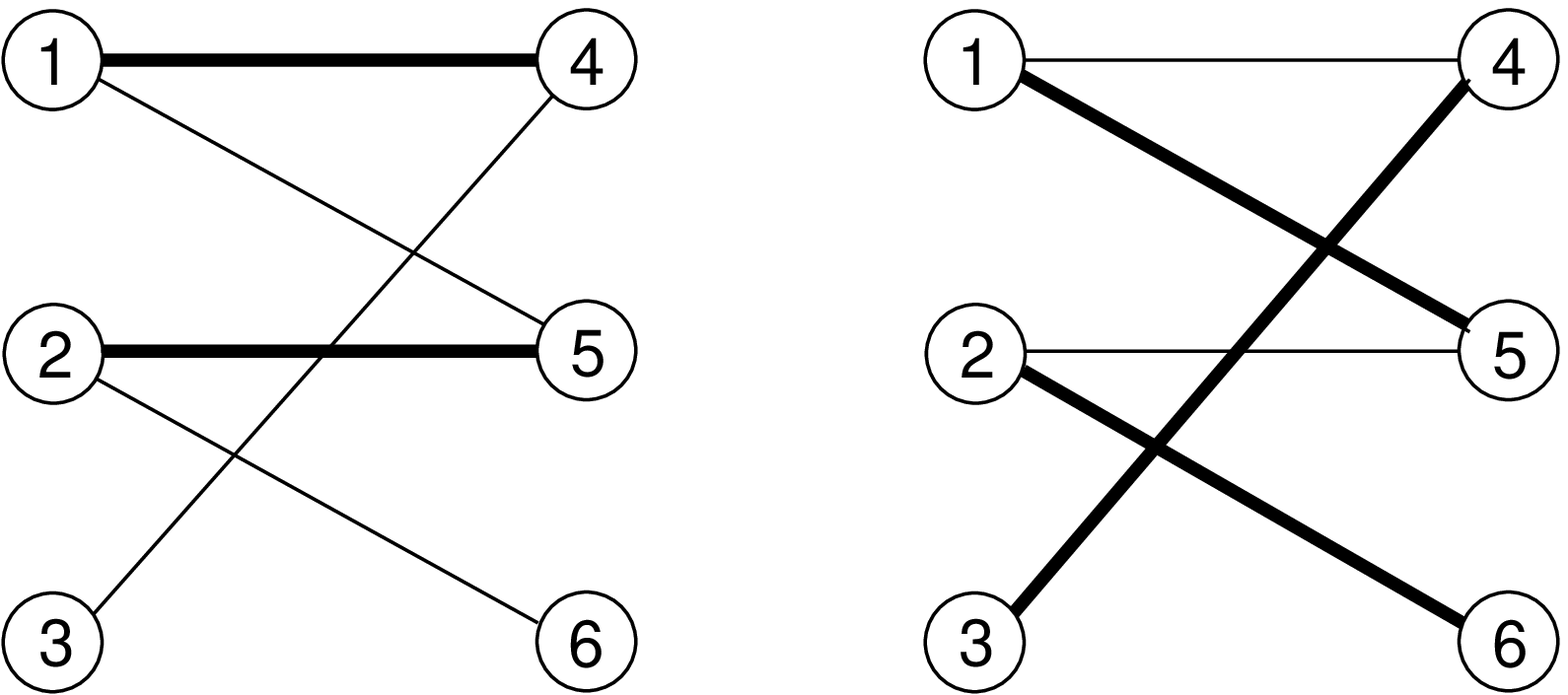}}
\end{center}
\narrowcaption{Example graph for matching, see text.}{fig:intro-matching}
\end{vchfigure}
In Fig.~\ref{fig:intro-matching} a sample graph is shown. Please note 
that the graph is bipartite
with vertex sets $U=\{1,2,3\}$ and $W=\{4,5,6\}$. Edges contained in
the matching
are indicated by thick lines. The matching shown in the
left half is $M=\{\{1,4\}, \{2,5\}\}$. This means, e.\,g., edge $\{1,4\}$ is
matched, while edge $\{3,4\}$ is free. Vertices
 1,2,4 and 5 are covered, while vertices 3 and 6 are exposed.

In the right half of the figure, a perfect matching 
$M=\{\{1,5\}, \{2,6\}, \{3,4\}\}$ is shown, i.\,e., there
are no exposed vertices. 
\end{example}

Having a weighted graph $G=(V,E,w)$, we consider also \emph{weighted
matchings}\index{weighted matching|ii}\index{matching!weighted},
with the weight of a matching given by the sum of the weights of all
matched edges. $M$ is a \emph{maximum-weight matching}\index{matching!maximum-weight}\index{maximum-weight matching} if its
total weight assumes a maximum with respect to all
possible matchings. For perfect matchings, there is a simple mapping
between maximum-weight matchings and minimum-weight matchings, namely:
let $\tilde{w}_{ij} = W_{\rm max} - w_{ij}$, where $w_{ij}$ is the
weight of edge $\{i,j\}$ and $W_{\rm max} > {\rm
max}_{\{i,j\}}(w_{ij})$.  A maximum perfect matching on
$\tilde{w}_{ij}$ is then a minimum perfect matching on $w_{ij}$.
  
A good historical introduction to matching problems,
the origins of which may be traced to the beginnings of
combinatorics, may be found in Lov\'asz and Plummer \cite{GRAPH-MatchTh}.  
Matching is directly related to statistical mechanics 
because the partition function \index{partition function}
for the two-dimensional Ising model on the square lattice can be
found by counting \emph{dimer coverings}\index{dimer covering}
(= perfect matchings)~\cite{GRAPH-MatchTh}. 
This is a graph \emph{enumeration}\index{enumeration problem}
\index{problem!enumeration}
problem rather than an \emph{optimization}\/ problem as 
considered here. As a general rule, graph enumeration 
problems are \emph{harder}\/ than graph-optimization problems.  

\begin{vchfigure}[htb]
\begin{center}
\scalebox{0.5}{\includegraphics{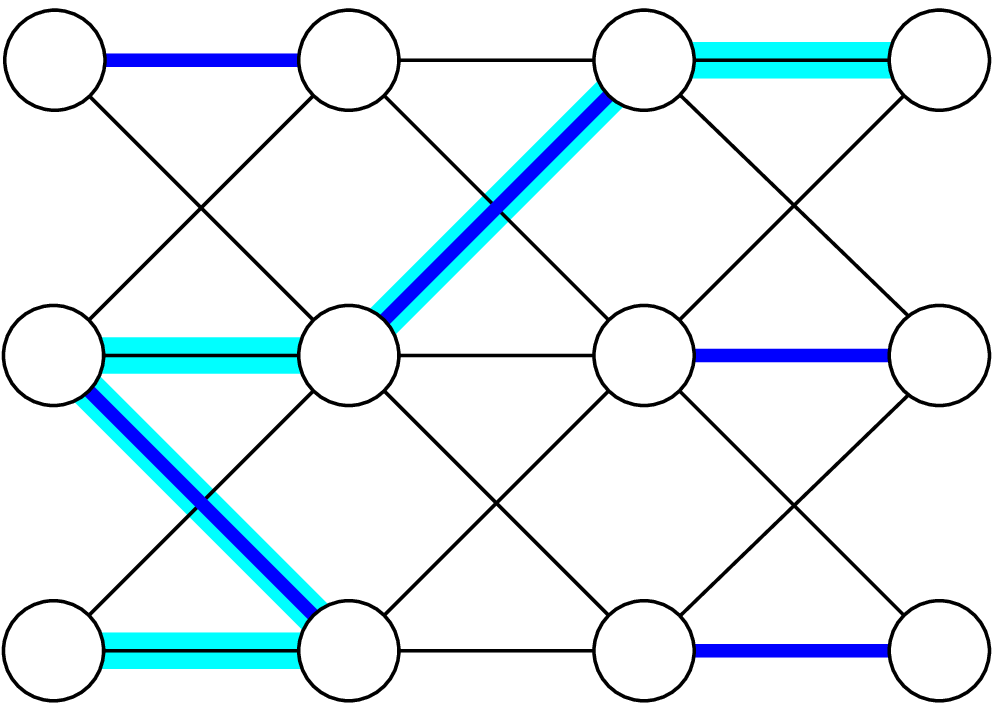}}
\quad\quad
\scalebox{0.5}{\includegraphics{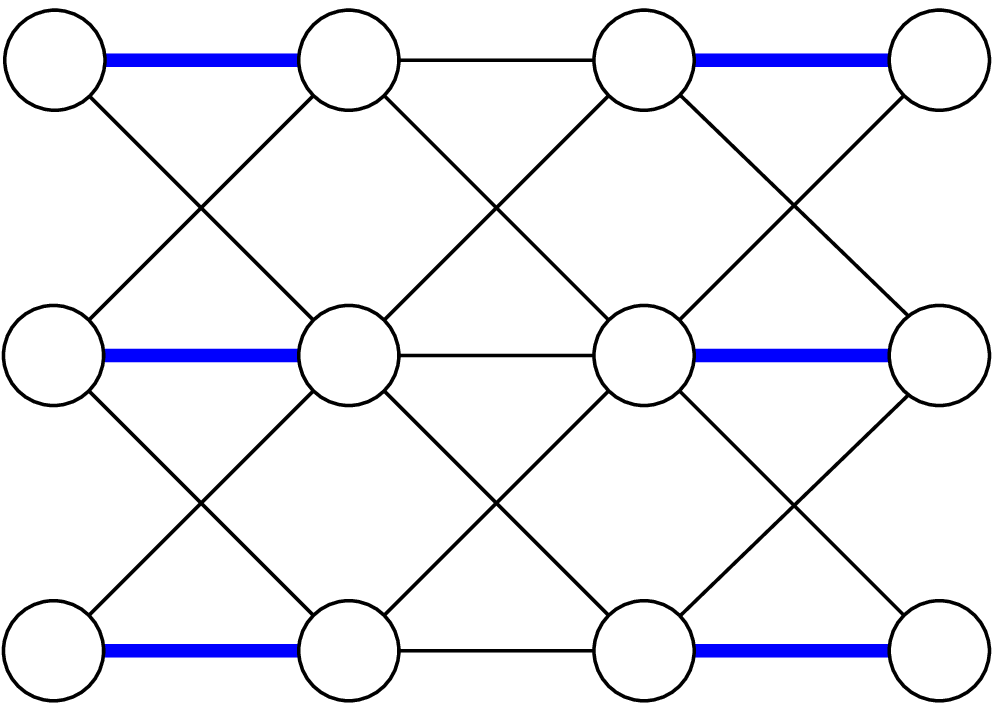}}
\vchcaption{{(Left):} A bipartite graph with a matching.
 An alternating path (shaded) starts and ends at exposed vertices, and
  alternates between unmatched (thin) and matched (thick) edges. {(Right):}
  Interchanging matched and unmatched edges along an alternating path increases
  the cardinality of the matching by one. This is called \emph{augmentation}
  and is the basic tool for maximum-matching algorithms. In this example,
  the augmentation yields a maximum-cardinality
  matching, even a perfect matching.}
\label{fig:matching-examples}
\end{center}
\end{vchfigure}

The ground state calculation of planar spin glasses is equivalent to the
minimum-weight perfect matching problem on a general graph, see
Sec.~11.7.  Maximum/minimum (perfect) matching problems on
general graphs are algorithmically more complicated than on bipartite graphs,
see, e.\,g., Refs~\cite{GRAPH-cook1998,GRAPH-korte2000}, but still they can
be solved in a running time increasing only polynomially with the system size.
These matching algorithms rely on finding \emph{alternating paths}
\index{alternating path}\index{path!alternating}, which are 
paths along
which the edges are alternately matched and unmatched, see
Fig.~\ref{fig:matching-examples}.

}

\section{Basic graph algorithms}
\label{sec:graph-algs}
\index{graph!algorithms|(}\index{algorithm!graph|(}

In this section, we present basic graph algorithms, which are related to
finding the connected components of a graph. First, we explain two different
strategies, the depth-first search and the breadth-first search. In addition
to yielding the connected components, they construct also spanning tress. Then
we explain how the strongly connected components of directed graphs can be
obtained by modifying the depth-first search. At the end we show, for the case
of weighted graphs, that minimum-weight spanning tress can be found using a
simple algorithm by Prim.  Another fundamental type of graph algorithm is the
shortest-path algorithm, which we explain later on in
Sec.~11.4.

\subsection{Depth-first and breadth-first search}
\label{subsec:depth-breadth}

\change{
For a given graph $G=(V,E)$, we want to determine its connected components.
There are two basic search
strategies, \emph{depth-first search} (DFS) 
\index{DFS|see{depth-first search}}
and
\emph{breadth-first search} (BFS), \index{BFS|see{breadth-first search}}
which are closely related.

\index{depth-first search|(}\index{algorithm!depth-first search|(}
The main work of the first algorithm is done within the following procedure
depth-first(), which starts at a given vertex $i$ and visits all
vertices which are connected to $i$. The main idea is to perform
recursive calls of depth-first() for all neighbors of $i$ which have
not been visited at that moment. The array \emph{comp}\/$[]$ is used to keep
track of the process. If \emph{comp}\/$[i]=0$ vertex $i$ has not been visited
yet. Otherwise it contains the number (id) $t$ of the component
currently explored. The value of $t$ is
also passed as a parameter, while
the array \emph{comp}\/$[]$ is passed as a reference, i.\,e., it behaves
like a global variable and all changes performed  to $comp[]$ in a
call to the procedure
persist after the call is finished.

\begin{procedure}{depth-first($G$, $i$, [ref.] \emph{comp}\/, $t$)}
\> \emph{comp}\/[$i$]:=$t$;\\
\> \textbf{for} all neighbors $j$ of $i$ \textbf{do}\\
\>\> \textbf{if} \emph{comp}\/[$j$] = 0 \textbf{then}\\
\>\>\> depth-first($G$, $j$, \emph{comp}\/, $t$);\\
\end{procedure}

To obtain all connected components, one has to call depth-first() for
all vertices which have not been visited yet. This is done by the
following algorithm. \index{component}\index{algorithm!components}

\begin{algorithm}{components($G$)}
\> initialize \emph{comp}\/$[i]:=0$  for all $i\in V$;\\
\> $t:=1$;\\
\> \textbf{while} there is a vertex $i$ with \emph{comp}\/$[i]$=0 \textbf{do}\\
\>\> depth-first($G$, $i$, \emph{comp}\/, $t$);\ \ $t:=t+1$;\\
\end{algorithm}

\index{spanning tree|(}\index{tree!spanning|(}
\index{depth-first spanning tree|(}
\index{spanning tree!depth-first|(}
For any vertex $i$, when the procedure calls itself recursively for a neighbor
$j$, it follows the edge $\{i,j\}$. Since no vertex is visited twice, those
edges form a spanning tree of each connected component. We call those edges
\emph{tree edges} and the other edges are known as \emph{back
  edges}. The algorithm 
tries to go 
as far as possible within a graph, before visiting further neighbors of the
vertex in which the search started. It is for this reason that the procedure
received its name.  Thus, the spanning tree is called the \emph{depth-first
  spanning tree} and the collection of the spanning trees of all connected
components is a depth-first spanning forest.  \index{spanning forest} In the
next section, we will use the depth-first spanning trees to find the strongly
connected components in a directed graph.

\begin{example}{Depth-first spanning tree}
  As an example, we consider the graph shown on the left of
  Fig.~\ref{fig:spanningA}. We assume that the algorithm first calls
  depth-first() for vertex 1, hence $comp[1]:=1$ is set.

\begin{vchfigure}[ht]
\begin{center}
  \scalebox{0.5}{\includegraphics{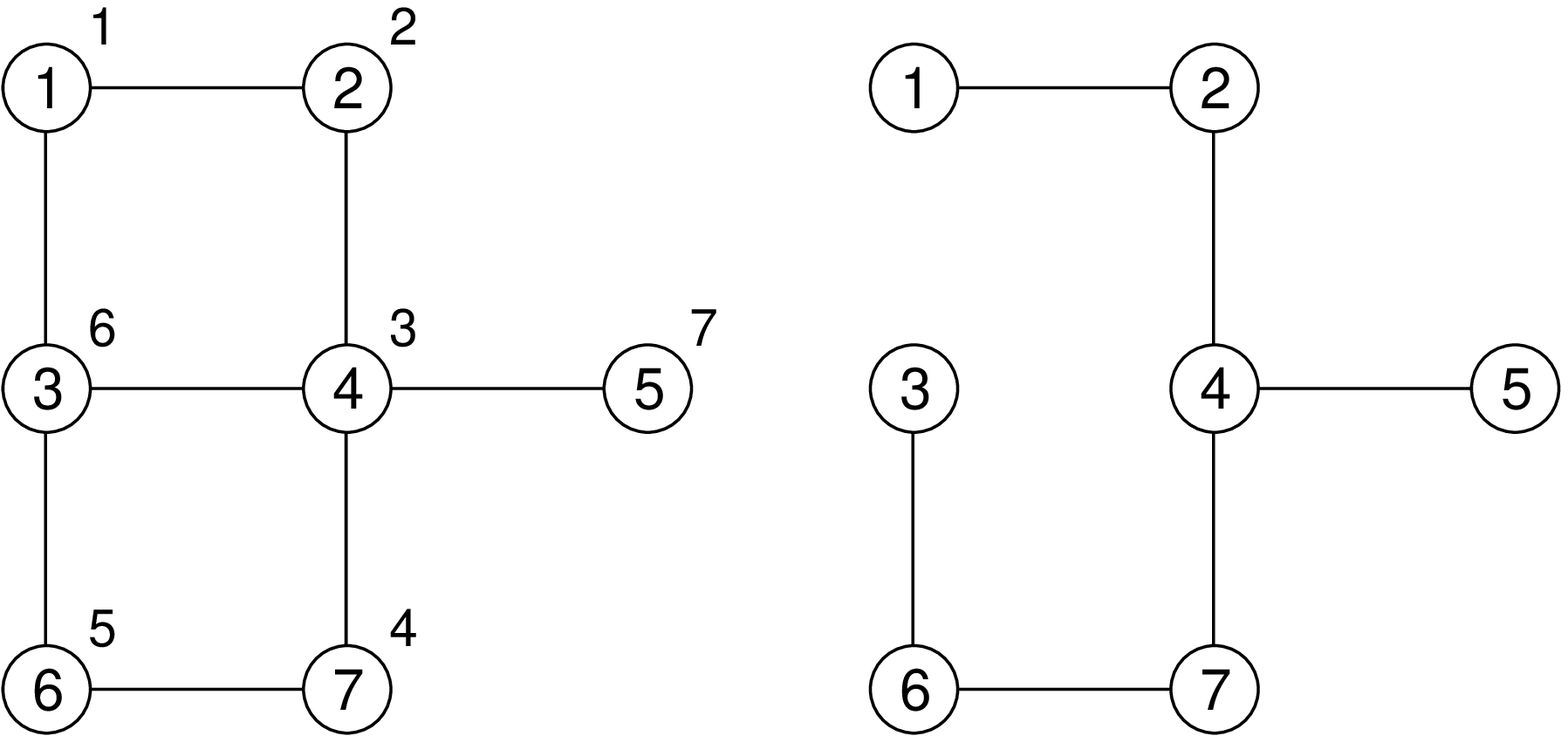}}
  \narrowcaption{A sample graph (left). The number close to the vertices
    indicates a possible order in which the vertices are visited during a
    depth-first search. On the right, the resulting depth-first spanning tree
    is shown.}{fig:spanningA}
\end{center}
\end{vchfigure}

We assume that vertex $2$ is the first neighbor of vertex 1
encountered in the \textbf{for} loop. Hence depth-first is called for
vertex $2$, where $comp[2]:=1$ is set. Here, in the \textbf{for} loop,
vertex 1 may be encountered first, but it has already been visited
($comp[1]=1$), hence nothing happens. In the next iteration the second
neighbor, vertex $4$ is encountered. Now depth-first() is called for
vertex $4$ and there $comp[4]:=1$ is assigned. We assume that vertex 7 is
the first vertex encountered in the loop over all neighbors of vertex
4. Therefore, depth-first($G,7, comp, 1$) is called next, 
leading to two more recursive
calls of depth-first(). The full recursive hierarchy of calls appears as
follows (we show only the parameter for vertex $i$):

\begin{tabbing} xx \= xx \= xx \= xx \= xx \= xx \= xx \=
  xxxxxxxxxxxxxxxxxxxxx  \= xxx \kill
\> depth-first($1$)\\
\>\> depth-first($2$)\\
\>\>\> depth-first($4$)\\
\>\>\>\> depth-first($7$)\\
\>\>\>\>\> depth-first($6$)\\
\>\>\>\>\>\> depth-first($3$)\\
\>\>\>\> depth-first($5$)
\end{tabbing}

The deepest point in the recursion is reached for vertex 3. All its neighbors,
vertices 1, 4 and 6 have already been visited, hence the procedure returns
here without further calls. At this point only vertex 5 has not been visited,
hence it is visited when iterating over the other neighbors of vertex 4.  The
order in which the vertices are visited is indicated in
Fig.~\ref{fig:spanningA} and the resulting depth-first spanning tree is also
shown.
\end{example}

During a run of the algorithm, for each vertex all neighbors are
accessed to test whether it has already been visited. 
Thus each edge is touched twice, which results in a time
complexity of ${\cal O}(|E|)$.

The depth-first search can also be adapted for directed graphs. 
\index{directed graph}\index{graph!directed} In this case,
being at vertex~$i$, one follows only edges ($i,j$), i.\,e., which point in a
forward direction. This means that the resulting depth-first spanning trees
may look different. Even the vertices contained in a tree may differ,
depending on the order in which the vertices are visited in the outer loop (as
in components() for the undirected case), as we will see in the example below.
This means that components are not well defined for directed graphs, because
they depend on the order in which the vertices are treated. Instead, one is
interested in calculating the strongly connected components, which are well
defined. The corresponding algorithm is presented in the next section. There
we will need the the \emph{reverse topological order}, 
\index{reverse topological order|ii}\index{topological order!reverse|ii}%
 also called
\emph{postorder}. \index{postorder} 
This is just the order in which the treatment of a vertex
$i$ \emph{finishes} during the calculation of the depth-first spanning tree,
i.\,e., the order in which the calls to the procedure, with $i$ as argument,
terminate.

The algorithm reads as follows, it is only slightly modified with respect to
the undirected case. The postorder is stored in array $post[]$, which is
passed as reference, i.\,e., all modifications to the values are also
effective outside the procedure. We also need a counter $c$, also passed by
reference, which is used to establish the postorder. The array $tree$, passed
also by reference, keeps track of which vertices have already been visited and
stores the identification numbers of the trees in which the vertices are
contained, corresponding to the array $comp$ used for the undirected case.

\begin{procedure}{depth-first-directed($G$, $i$, [ref.] $tree$, $t$, 
[ref.] $post$, [ref.] $c$)}
\> $tree[i]:=t$;\\
\> \textbf{for} all $j$ with $(i,j)\in E$  \textbf{do}\\
\>\> \textbf{if} \emph{tree}\/[$j$] = 0 \textbf{then}\\
\>\>\> depth-first-directed($G$, $j$, $tree$, $t$, $post$, c);\\
\> $post[i]:=c$;\\
\> $c:=c+1$; \\
\end{procedure}

The main subroutine, finding all connected depth-first spanning
trees and establishing the reverse topological order, reads
as follows.
\newpage
\begin{algorithm}{postorder($G$)}
\> initialize \emph{tree}\/$[i]:=0$  for all $i\in V$;\\
\> $t:=1$;\\
\> $c:=1$;\\
\> \textbf{while} there is a vertex $i$ with $tree[i]$=0 \textbf{do}\\
\>\> depth-first-directed($G$, $i$, $tree$, $t$, $post$, $c$);
      t:=t+1; \\
\end{algorithm}

The vertices, for which depth-first-directed() is called
inside the loop in postorder(), i.\,e., the vertices where the
construction of a tree starts, are called \emph{roots} of the
depth-first spanning trees. They are the vertices which receive the
highest $post$ numbers for each tree. Vertices of other trees, 
which are visited earlier
receive lower $post$ numbers, vertices of trees
being visited later will receive higher $post$ numbers than all vertices
of the current tree.

\begin{example}{Depth-first spanning tree of a directed graph}
We consider the graph shown in Fig.~\ref{fig:directedDFSA}.
\begin{vchfigure}[ht]
\begin{center}
\scalebox{0.5}{\includegraphics{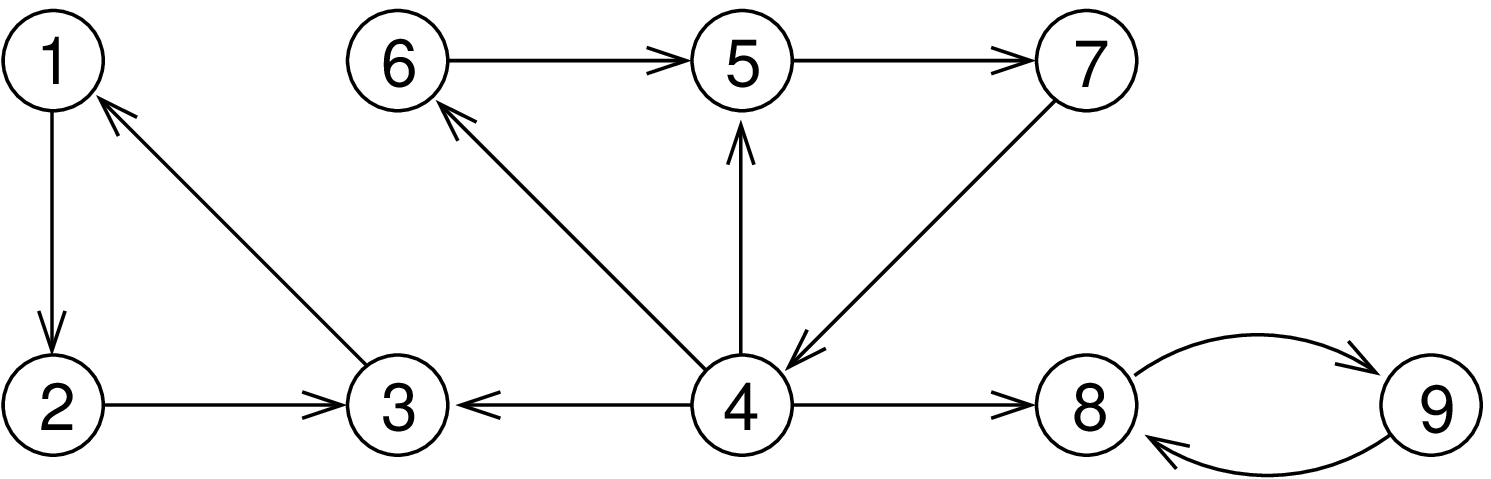}}
\narrowcaption{A sample directed graph.}{fig:directedDFSA}
\end{center}
\end{vchfigure}

We assume that in postorder()  vertex 1 is first considered. Vertex 1
has one neighbor, hence depth-first-directed() will be called for vertex
2, in which the procedure is called for vertex 3. Vertex 3 has one
neighbor in the forward direction, vertex 1, which has already been
visited, hence no call is performed here. This means that the procedure
finishes for vertex 3, and $post[3]:=1$ is assigned. Since all neighbors 
of vertex 2 have been visited,  this call also finishes and $post[2]=2$ is
assigned. Next the same thing happens for vertex 1, resulting in $post[1]=3$.
Hence, the algorithm returns to the top level, to postorder().

Now, we assume that depth-first-directed is called for vertex 4, and that
its neighbors are processed in the order 3,5,6,8. Vertex 3 has already been
visited. The call of depth-first-directed() for vertex 5 leads to a call of
the procedure for its unvisited neighbor vertex 7. The call
for vertex 8 itself calls the procedure for vertex 9. In total the following
calling hierarchy is obtained, only the passed values for the vertex $i$ 
are shown.\newpage
\label{page:dfs-run1}

\begin{tabbing} xx \= xx \= xx \= xx \= xx \= xx \= xx \=
  xxxxxxxxxxxxxxxxxxxxx  \= xxx \kill
\> depth-first-directed($1$)\\
\>\> depth-first-directed($2$)\\
\>\>\> depth-first-directed($3$)\\
\> depth-first-directed($4$)\\
\>\> depth-first-directed($5$)\\
\>\>\> depth-first-directed($7$)\\
\>\> depth-first-directed($6$)\\
\>\> depth-first-directed($8$)\\
\>\>\> depth-first-directed($9$)\\
\end{tabbing}

The resulting depth-first spanning forest is shown in
Fig.~\ref{fig:directedDFSB} together with the $post$ values for the reverse
topological order. The roots of the trees are always shown at the top.
\begin{vchfigure}[ht]
\begin{center}
\scalebox{0.5}{\includegraphics{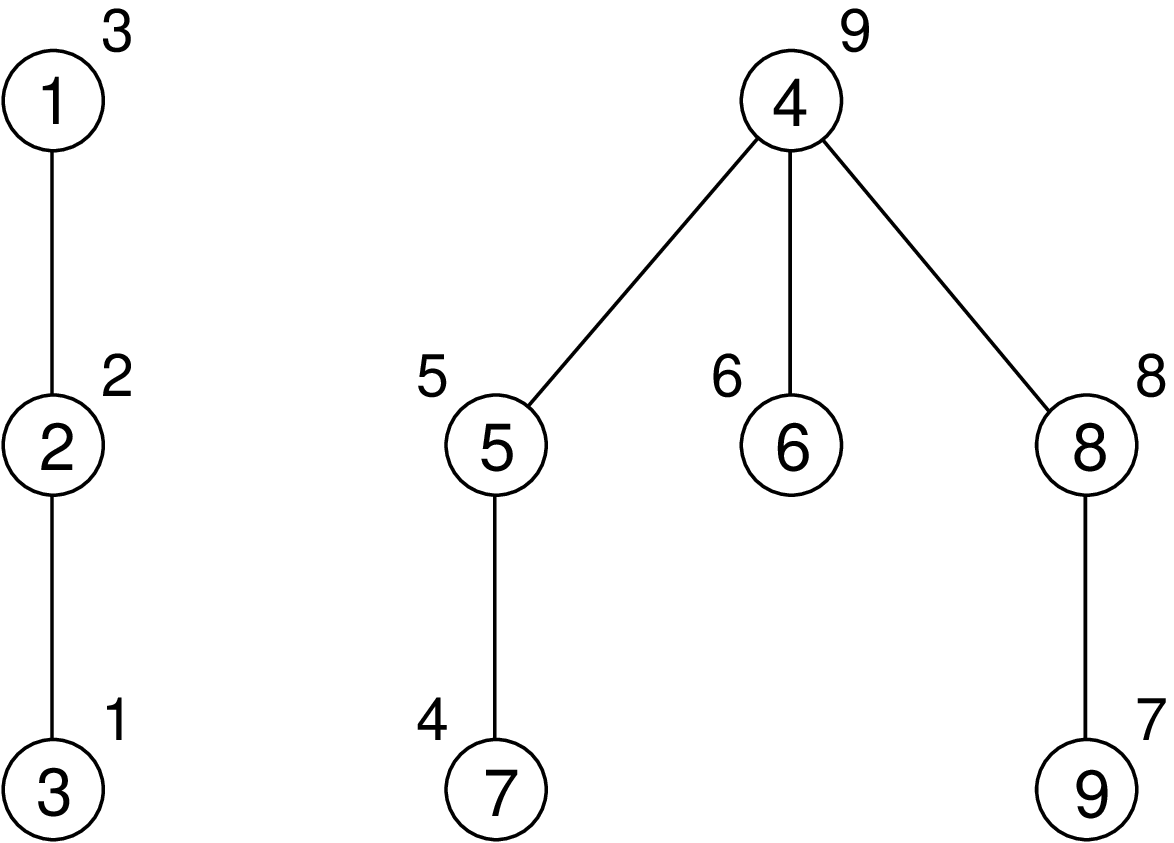}}
\narrowcaption{On a possible depth-first spanning forest of the graph
from Fig.~\ref{fig:directedDFSA}. The numbers close to the vertices
denote the $post$ values of the reverse topological order.}{fig:directedDFSB}
\end{center}
\end{vchfigure}

If the vertices are treated in a different order, then the resulting
depth-first spanning forest might look different. 
When, e.\,g., vertex 4 is considered first,
then all vertices will be collected in one tree. The resulting tree,
assuming that the neighbors of vertex 4 are considered in the order 8,6,5,3,
is shown in Fig.~\ref{fig:directedDFSC}.
\begin{vchfigure}[ht]
\begin{center} 
\scalebox{0.5}{\includegraphics{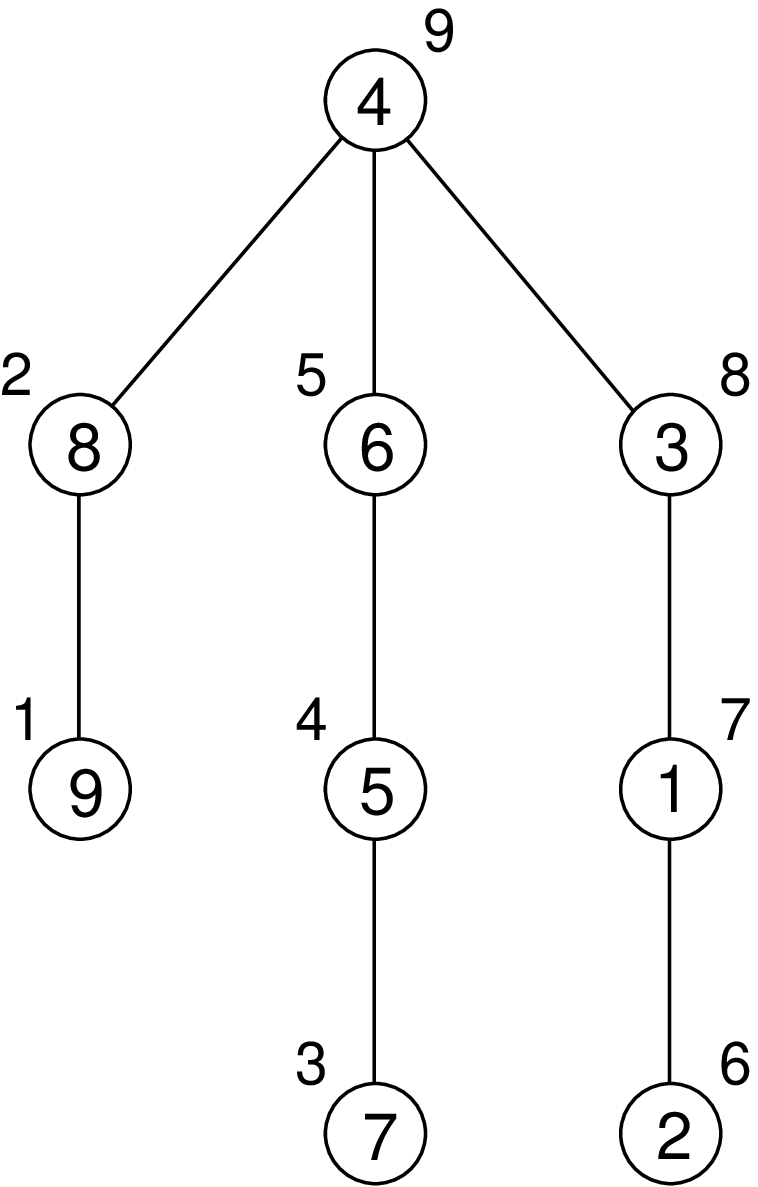}}
\narrowcaption{Another possible depth-first spanning tree of the graph
from Fig.~\ref{fig:directedDFSA}. The numbers close to the vertices
denote the $post$ values of the reverse topological order.}{fig:directedDFSC}
\end{center}
\end{vchfigure}
\end{example}
\index{depth-first spanning tree|)}
\index{spanning tree!depth-first|)}
\index{depth-first search|)}\index{algorithm!depth-first search|)}

\index{breadth-first spanning tree|(}
\index{spanning tree!breadth-first|(}
\index{breadth-first search|(}\index{algorithm!breadth-first search|(}
Now we turn back to undirected graphs. 
A similar algorithm to
a depth-first search is an algorithm, which first visits all neighbors of a
vertex before proceeding with vertices further away. This algorithm is called
a \emph{breadth-first search}\/.
 This means that at first
all neighbors of the initial vertex are visited.  This initial vertex is the
root of the \emph{breadth-first spanning tree} being built.  These neighbors
have distance one from the root, when measured in terms of the number of edges
along the shortest path from the root.  In the previous example in
Fig.~\ref{fig:spanningA}, the edge $\{1,2\}$ would be included in the spanning
tree, if it is constructed using a breadth-first search, as we will see below.
In the next step of the algorithm, all neighbors of the vertices treated in
the first step are visited, and so on. Thus, a queue\footnote{A queue is a
  linear list, where one adds elements on one side and removes them from the
  other side.} can be used to store the vertices which are to be processed.
The neighbors of the current vertex are always stored at the end of the queue.
Initially the queue contains only the root. The algorithmic representation
reads as follows, $level(i)$ denotes the distance of vertex $i$ from the root
$r$ and $pred(i)$ is the predecessor of $i$ in a shortest path from $r$, which
defines the {breadth-first spanning tree}.

\begin{algorithm}{breadth-first search($G,r$, [ref.] $comp,t $)}
\> Initialize queue $Q:=\{r\}$;\\
\> Initialize $level[r]:=0$; $level[i]:=-1$ (undefined) 
for all other vertices;\\
\> $comp[r]:=t$;\\
\> Initialize $pred[r]:=-1$;\\
\> \textbf{while} $Q$ not empty \\
\> \textbf{begin}\\
\>\> Remove first vertex $i$ of $Q$;\\
\>\> \textbf{for} all neighbors $j$ of $i$ \textbf{do}\\
\>\>\> \textbf{if} $level[j]=-1$ \textbf{then}\\
\>\>\> \textbf{begin}\\
\>\>\>\> $level[j] := level[i]+1$; \\
\>\>\>\> $comp[j] := t$;\\
\>\>\>\> $pred[j]:=i$;\\
\>\>\>\> add $j$ at the end of $Q$;\\
\>\>\> \textbf{end}\\
\> \textbf{end}\\
\end{algorithm}
Also for a breadth-first search, for each vertex all neighbors are
visited. Thus each edge is again touched twice, which results in a time
complexity of ${\cal O}(|E|)$. To obtain the breadth-first 
spanning forest of a 
graph, one has to call the procedure for all yet unvisited vertices inside a
loop over all vertices, as in the algorithm components() presented above.

\begin{example}{Breadth-first search}
  We consider the same graph as in the example above, shown now in Fig.
\ref{fig:spanningB}. Initially the queue contains the source,
here 1 again and
all values $level[i]$ are ``undefined'' (-1), except $level[1]=0$.

$Q=\{0\}$, $level[1]=1$

\begin{vchfigure}[ht]
\begin{center}
  \scalebox{0.5}{\includegraphics{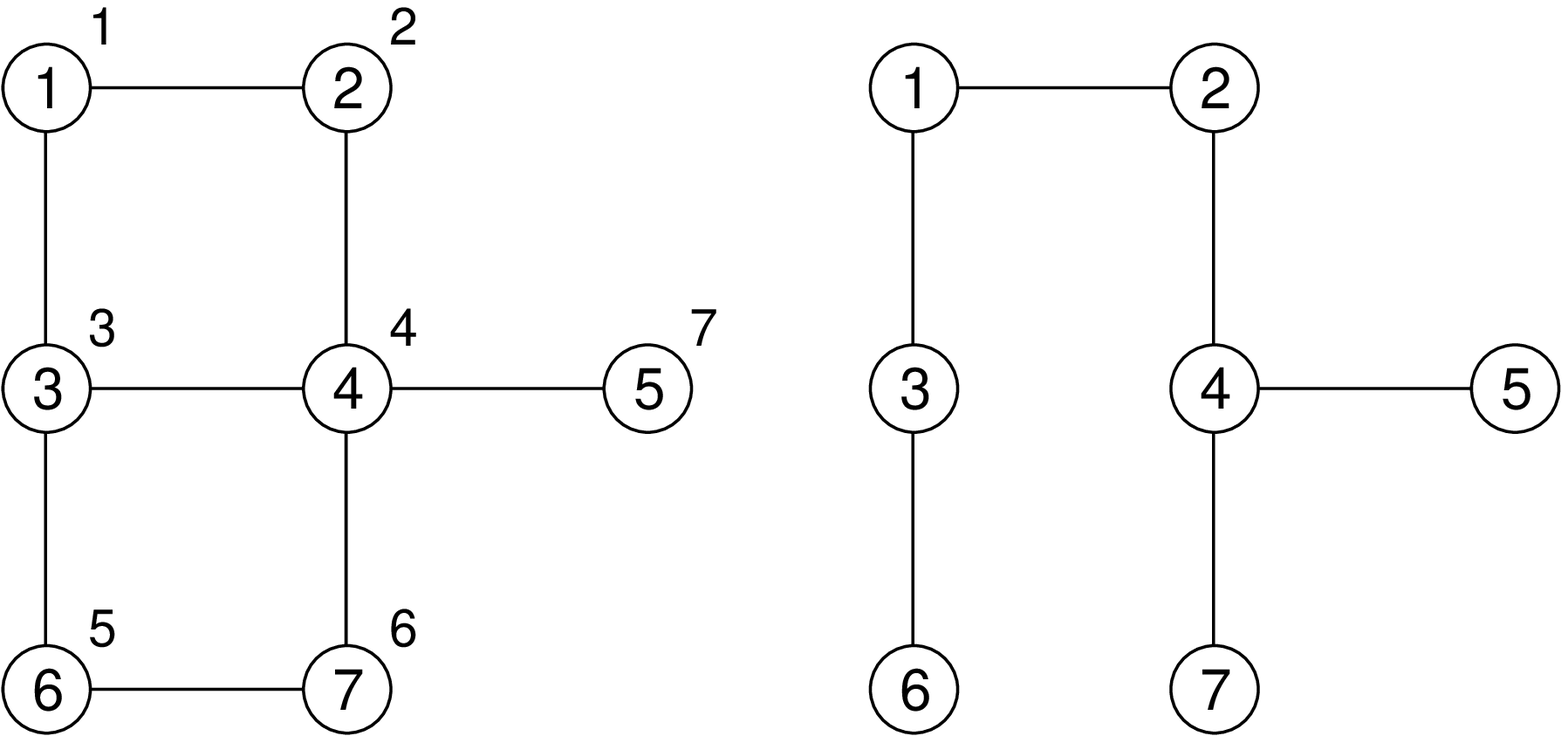}}
  \narrowcaption{A sample graph (left). The number close to the vertices
    indicate a possible order in which the vertices are visited during a
    breadth-first search. On the right the resulting breadth-first spanning
    tree is shown.}{fig:spanningB}
\end{center}
\end{vchfigure}

While treating vertex 1, its neighbors, vertices 2 and 3, are added
to the queue, thus $pred[2]=pred[3]=1$. They have a distance 1 from the source
($level[2]=level[3]=1$).

$Q=\{2,3\}$, $level[1]=0$, $level[2]=level[3]=1\,.$

Next vertex 2 is processed. It has two neighbors, vertices 1 and 4, but
vertex 1 has been visited already $(level[1] \neq -1$), thus only
vertex 4 is added to $Q$ with $pred[4]=2$, $level[4]=level[2]+1=2$.
After this iteration the situation is as follows:

$Q=\{3,4\}$, $level[1]=0$, $level[2]=level[3]=1$, $level[4]=2\,.$

The treatment of vertex 3 adds vertex 6 to the queue
($level[6]=level[3]+1=2$, $pred[6]=3$). At the beginning of the next
iteration vertex 4 is taken.  Among its four neighbors, vertices 5 and
7 have not been visited.
 Thus $level[5]=level[7]=3$ and $pred[5]=pred[7]=4$. Now
all values of the $pred$ and $level$ 
arrays are set. Finally, vertices 6,5 and
7 are processed, without any change in the variables. 

The resulting breadth-first spanning tree is shown on the right of
Fig.~\ref{fig:spanningB}. It is also stored in the array $pred$ of shortest
paths to the source, e.\,g., a shortest path from vertex 7 to vertex 1 is given
by the iterated sequence of predecessors: $pred[7]=4, pred[4]=2, pred[2]=1$.
\end{example}
\index{breadth-first search|)}\index{algorithm!breadth-first search|)}
\index{breadth-first spanning tree|)}
\index{spanning tree!breadth-first|)}
\index{spanning tree|)}\index{tree!spanning|)}

}

\subsection{Strongly connected component}
\label{subsec:scc}
\index{strongly connected component|(}
\index{algorithm!strongly connected component|(}

\change{
As we recall from Sec.~\ref{subsec:graph-defs}, the strongly connected 
components of a directed graph $G=(V,E)$ 
are the maximal subsets of vertices in which,
for each pair $i,j$, a directed path from $i$ to $j$ and a directed path
from $j$ to $i$, exists in $G$. As we will show here, they can be obtained
by \emph{two} depth-first searches. \index{depth-first search}
\index{algorithm!depth-first search}

The basic idea is as follows. First one performs a depth-first search on $G$, 
obtaining the depth-first spanning forest and the 
reverse topological order in the array $post$. Next,
once constructs the \emph{reverse} graph \index{reverse graph|ii}
\index{graph!reverse|ii} $G^R$, which is obtained
by using the same vertices as in $G$ and with all edges from $G$ reversed
\begin{equation}
G^R\equiv(V,E^R)\quad E^R\equiv\{(j,i)|(i,j)\in E\}\,.
\end{equation} 
Then we compute the depth-first spanning forest of $G^R$, but \emph{we treat
  the vertices in the main loop in decreasing values of the reverse
  topological order} \index{reverse topological order}
\index{topological order!reverse} obtained in the 
first computation of the depth-first
order of $G$.  In particular, we always start with the vertex having received
the highest $post$ number. The trees of the depth-first spanning forest
obtained in this way are the strongly connected components of $G$.  The
reason is that, for each tree, the vertices with highest $post$ number are
those from where all vertices of a depth-first tree can be reached, i.\,e.,
the \emph{roots} \index{root}\index{tree!root}
of the trees.  When starting the depth-first search at these
vertices for the reverse graph, graph \index{reverse graph}
\index{graph!reverse} one is able to visit the vertices, from
which one is able to reach the root in the original graph. In total these are
the vertices from which one can reach the root and which can be reached
starting from the root, i.\,e., they compose the strongly connected
components.  A detailed proof will be given below.
The algorithm can be summarized as follows:
\begin{algorithm}{scc($G$)}
\> initialize \emph{tree}\/$[i]:=0$  for all $i\in V$;\\
\> $t:=1$; $c:=1$;\\
\> \textbf{while} there is a vertex $i$ with $tree[i]$=0 \textbf{do}\\
\>\> depth-first-directed($G$, $i$, $tree$, $t$, $post1$, $c$);
t:=t+1\\
\> calculate reverse graph $G^R$;\\
\> initialize \emph{tree}\/$[i]:=0$  for all $i\in V$;\\
\> $t:=1$;\\
\> $c:=1$;\\
\> \textbf{for} all $i\in V$ in decreasing value of $post1[i]$ \textbf{do}\\
\>\> \textbf{if} $tree[i]$=0 \textbf{do}\\
\>\>\> depth-first-directed($G^R$, $i$, $tree$, $t$, $post2$, $c$);
        t:=t+1;\\
\end{algorithm}

\begin{example}{Strongly connected component}
As an example, we consider the graph from Fig.~\ref{fig:directedDFSA}. 
We assume that the reverse topological order 
\index{reverse topological order}%
\index{topological order!reverse} is as obtained in the
first example run 
which was shown on page \pageref{page:dfs-run1}, 
hence the vertices ordered in decreasing
$post$ numbers of the reverse topological order are
\begin{equation*}
4,8,9,6,5,7,1,2,3
\end{equation*}
In this order, the depth-first search is applied to the reverse graph $G^R$,
graph \index{reverse graph}%
\index{graph!reverse} which is shown in Fig.~\ref{fig:sccA}.

\begin{vchfigure}[ht]
\begin{center}
\scalebox{0.5}{\includegraphics{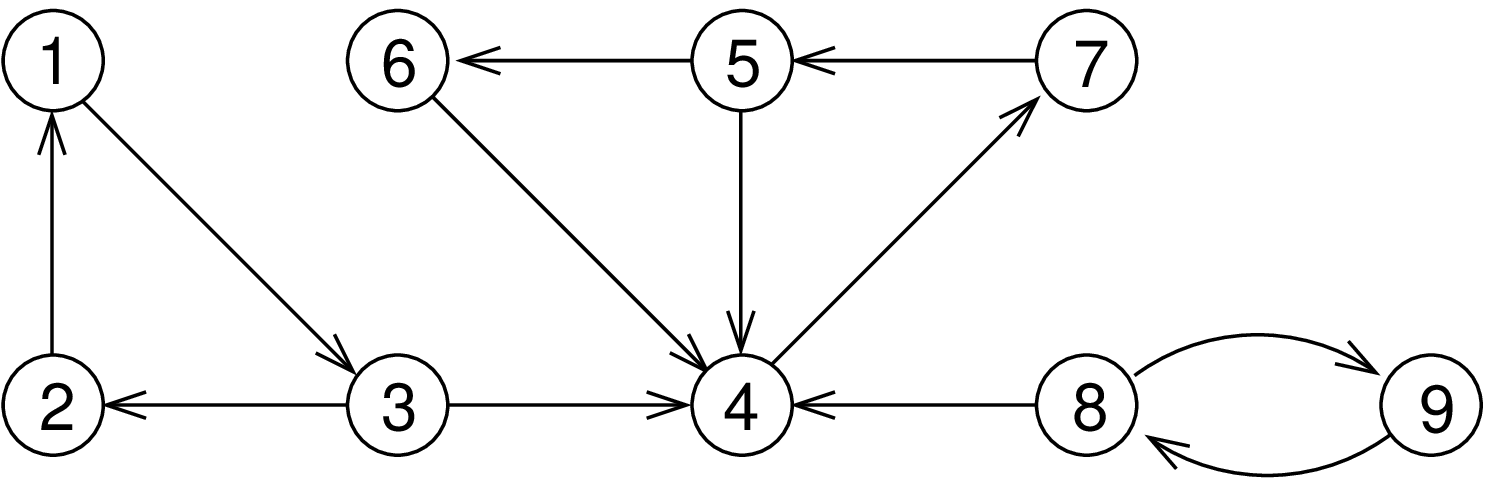}}
\narrowcaption{The reverse graph of graph shown in Fig.~\ref{fig:directedDFSA}.}{fig:sccA}
\end{center}
\end{vchfigure}

\begin{minipage}[t]{0.5\textwidth}
Hence, when calling depth-first-directed($G^R$, 4, $tree$, $1$, $post2$, $c$),
the vertices 7, 5, and 6 are visited recursively in this order, before the
call with $i=4$ terminates. Next, in the main loop with $t=2$,
the procedure is called
for vertex $i=8$, which has the second highest value in $post1$, the vertices
$8$ and $9$ are processed. For the third ($t=3$) iteration,
depth-first-directed() is called for vertex 1, resulting in the visits
of also vertices 3 and 2. The resulting depth-first spanning forest
is shown on the right.

This means that
 we have obtained three trees, corresponding to the three strongly 
connected components $C_1=\{1,2,3\}$, $C_2=\{4,5,6,7\}$ and $C_2=\{8,9\}$.
\end{minipage}
\begin{minipage}[t]{0.5\textwidth}
\begin{center}
\rule{1pt}{0pt}

\scalebox{0.5}{\includegraphics{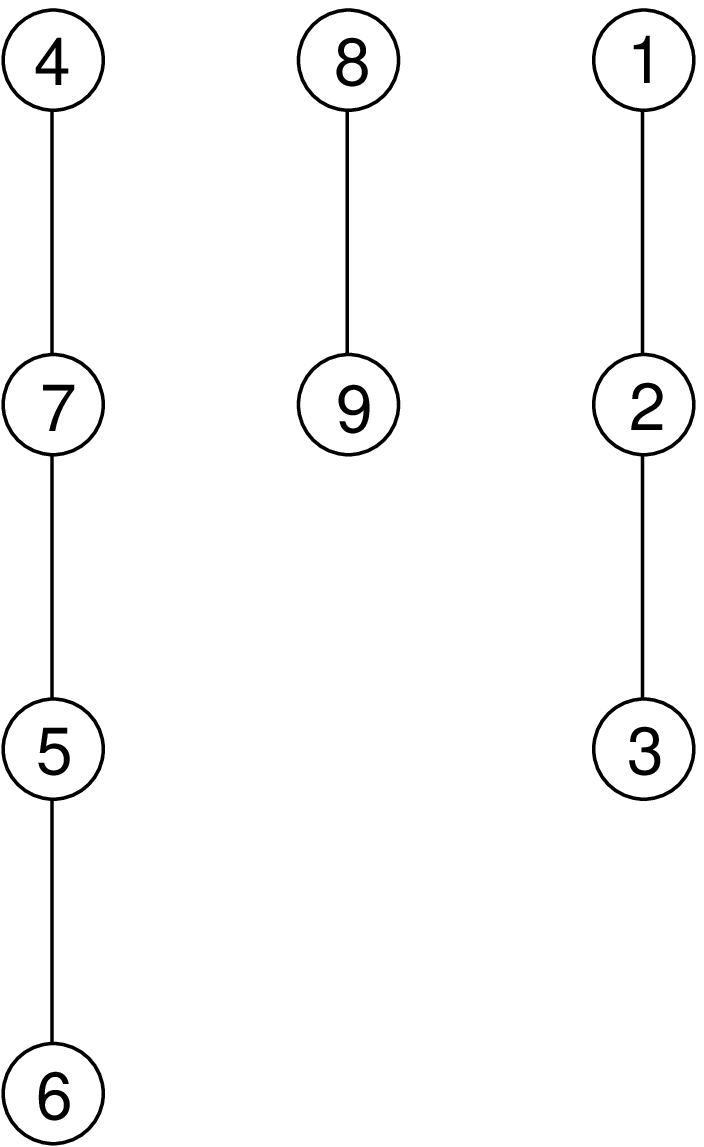}}
\end{center}
\end{minipage}
\vspace{-\baselineskip}

\end{example}

The algorithm consist of two loops over vertices, containing 
calls to depth-first-directed(). Each loop
takes time ${\cal O}(|V|+|E|)$, see Sec.~\ref{subsec:depth-breadth}. 
Note that one can obtain the vertices in decreasing reverse topological
order, by writing each vertex in an array after its call to
depth-first-directed() has finished. For brevity, we have not included
this array in the above algorithm. This means that 
one can perform the \textbf{for} loop in scc() simply by going through the
array in reverse order, i.\,e., without additional time cost.
Since also calculating the reverse graph graph \index{reverse graph}
\index{graph!reverse} can be done in ${\cal O}(|V|+|E|)$, the
full algorithm has a running time ${\cal O}(|V|+|E|)$. There are variants of
the algorithm, which require only one calculation of the depth-first
spanning forest instead of two, also no calculation of the reverse
graph occurs there. Instead, they require updating of additional data fields,
hence they are a bit harder to understand. The reader interested in these
algorithms should consult the standard literature 
\cite{GRAPH-aho74,GRAPH-sedgewick90,GRAPH-cormen2001}.

We close this section by showing that the above algorithm does indeed yield
the strongly connected components of a directed graph.  Note that the strongly
connected components of the reverse graph graph \index{reverse graph}
\index{graph!reverse} are exactly the same as those of the
graph itself.

\begin{proof}
$(\rightarrow)$\\
Assume that two vertices $i$ and $j$ are mutually reachable, i.\,e., are connected
by paths. Hence, they are mutually reachable in the reverse graph. This means
that during the second calculation of the forest, they will be in the
same tree.

\begin{minipage}[t]{0.54\textwidth}
$(\leftarrow)$

Now we assume that $i$ and $j$ are in the same tree of the 
depth-first forest calculation of $G^R$. Let $r$ be the root of this 
tree, i.\,e., the vertex for which depth-first-directed() has been called
on the top level in the \textbf{for} loop.  
Since the \textbf{for} loop is performed in decreasing order of
the $post$ values, this means that $r$ has the highest $post$ value
of the tree.
This means in particular $post[r]>post[i]$ and $post[r]>post[j]$ (*).

\end{minipage}
\quad
\begin{minipage}[t]{0.41\textwidth}
\rule{1pt}{0pt}

{\includegraphics[width=\textwidth]{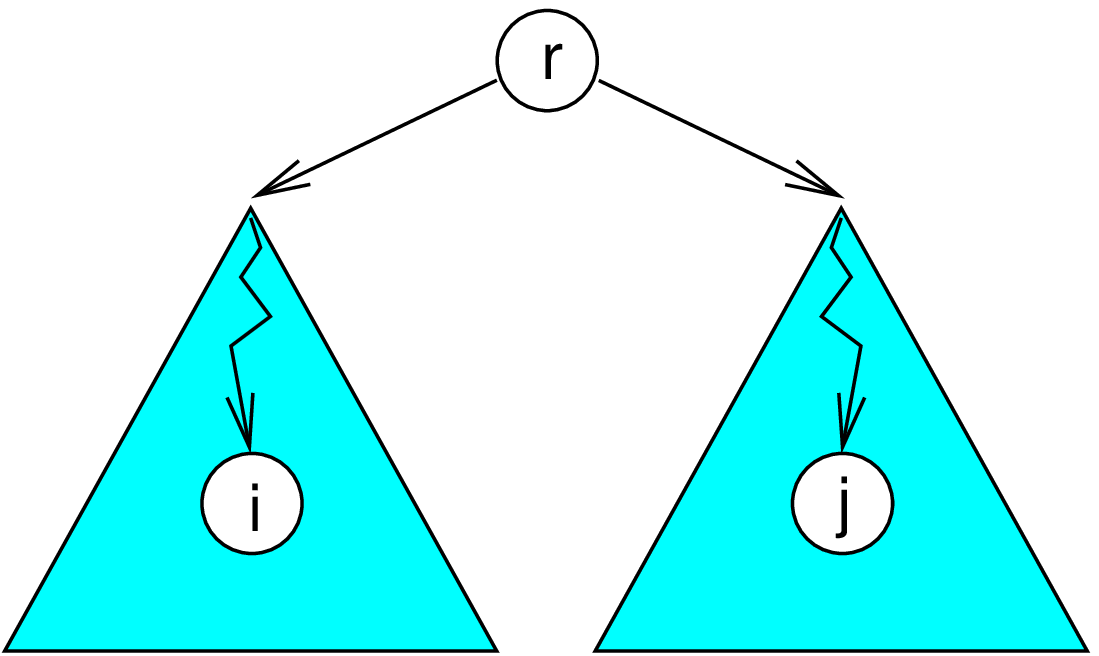}}
\end{minipage}

We perform the proof by raising a contradiction.  We assume 
that $i$ and $j$ are \emph{not} in the same strongly connected  component. 

Since $i$ and $j$  are in the same tree with root $r$ for $G^R$, 
there must paths from the root $r$ to $i$ and from $r$ to $j$
 in the reverse graph  graph \index{reverse graph}
\index{graph!reverse} $G^R$, hence there must be paths
from $i$ to $r$ and from $j$ to $r$ in $G$. 
Since $i$ and $j$ are assumed not to be  in the same
strongly connected component, either there is no path from $r$ to $i$
in $G$ or there is no path from $r$ to $j$ in $G$.

Without loss of generality, we consider the first case  $r\not\to i$.
There are two possibilities:
\begin{itemize}
\item[a)] $i$ is visited before $r$ in the calculation of the
depth-first forest of $G$. But then, because there is a path from $i$
to $r$ in $G$, the call to depth-first-directed() 
for $r$ would finish before the call for $i$, hence
we would have $post[i]>post[r]$. This is a contradiction to (*)!

\item[b)] $r$ is visited before $i$. Since there is no path from $r$
to $i$, vertex $i$ will  still be unvisited, when the call for $r$
has been finished, hence we will have again $post[i]>post[r]$.
This is a contradiction to (*)!
\end{itemize}

In the same way, we can raise a contradiction when assuming that path $r\to j$
does not exist in $G$. This means in total that $i$ and $j$ must be in the
same strongly-connected component.
\end{proof}

}
\index{strongly connected component|)}
\index{algorithm!strongly connected component|)}

\subsection{Minimum spanning tree}
\label{subsec:prims}
\index{Prim's algorithm|(}\index{algorithm!Prim's|(}
\index{minimum spanning tree|(}\index{spanning tree!minimum|(}

The algorithm used to construct a minimum spanning tree of a given graph 
\change{(see Sec.~\ref{subsec:spanning} for a definition)}
is \emph{greedy}. \index{greedy algorithm}\index{algorithm!greedy}
This means that at every
step the algorithm takes  the least expensive decision. In contrast to the
TSP-algorithm presented in Chap.~2, a global minimum is
guaranteed to be found.  \change{The basic idea of Prim's algorithm is that}
one starts with the edge of minimal weight,
and goes on \change{adding} minimal edges connected to the already constructed
subtree, unless a cycle would result:

\begin{algorithm}{Prim(G)}
  \> choose $\{i,j\}$ :
  $\omega_{ij} :={\rm  min}\{\omega_{km}| \{k,m\}\in E\}$;\\
  \> $S := \{i,j\}$;\\
  \> $\overline{S}:=V \setminus \{i,j\}$;\\
  \> $T := \left\{\{\s, r\}\right\}$;\\
  \> $X := \omega_{ij}$;\\
  \> \textbf{while} $|S|< |V|$ \textbf{do}\\
  \> \textbf{begin}\\
  \>\> choose $\{i,j\}$ : $i\in S,\ j\in \overline{S}$ and 
       $\omega_{ij} := {\rm min}\{\omega_{km}| k\in S,
       m\in\overline{S}, \{k,m\}\in E \}$;\\
  \>\> $S:=S\cup \{j\}$;\\
  \>\> $\overline{S}:=\overline{S} \backslash \{j\}$;\\
  \>\>  $T:=T\cup\left\{\{i,j\}\right\}$;\\
  \>\> $X:=X + \omega_{ij}$;\\
  \> \textbf{end}\\
  \> \textbf{return} $(S,T)$;\\
\end{algorithm}

The action of Prim's algorithm is illustrated in Fig.~\ref{fig:prim}.
One can easily verify that it produces in fact a minimum spanning
tree: Imagine, that this is not the case, i.\,e., in the algorithm you
have, for the first time, added a wrong edge $\tilde e$ at the $k$th
step. Now imagine that you have a minimum spanning tree coinciding with
the tree constructed in Prim's algorithm in the first $k-1$ selected
edges, but not containing $\tilde e$. If you add edge $\tilde e$ to
the minimum spanning tree, you introduce a cycle, which, by Prim's
algorithm, contains at least one edge of higher weight. Deleting the
latter thus gives a new spanning tree of smaller weight, which is a
contradiction to the minimality of the initial spanning tree.

Since each edge is chosen at most one, the while loop is performed ${\cal
  O}(|E|)$ times. In a simple version, choosing the edge with minimum weight
requires a loop over all edges (${\cal O}(|E|)$), while a more refined version
could use a \emph{priority queue} \index{priority queue}
\cite{GRAPH-aho74,GRAPH-sedgewick90,GRAPH-cormen2001}, such that the choose
operation takes only ${\cal O}(\log|E|)$ time, leading to a total of ${\cal
  O}(|E|\log|E|)$ running time.

\begin{vchfigure}[ht]
\begin{center}
\scalebox{0.7}{\includegraphics{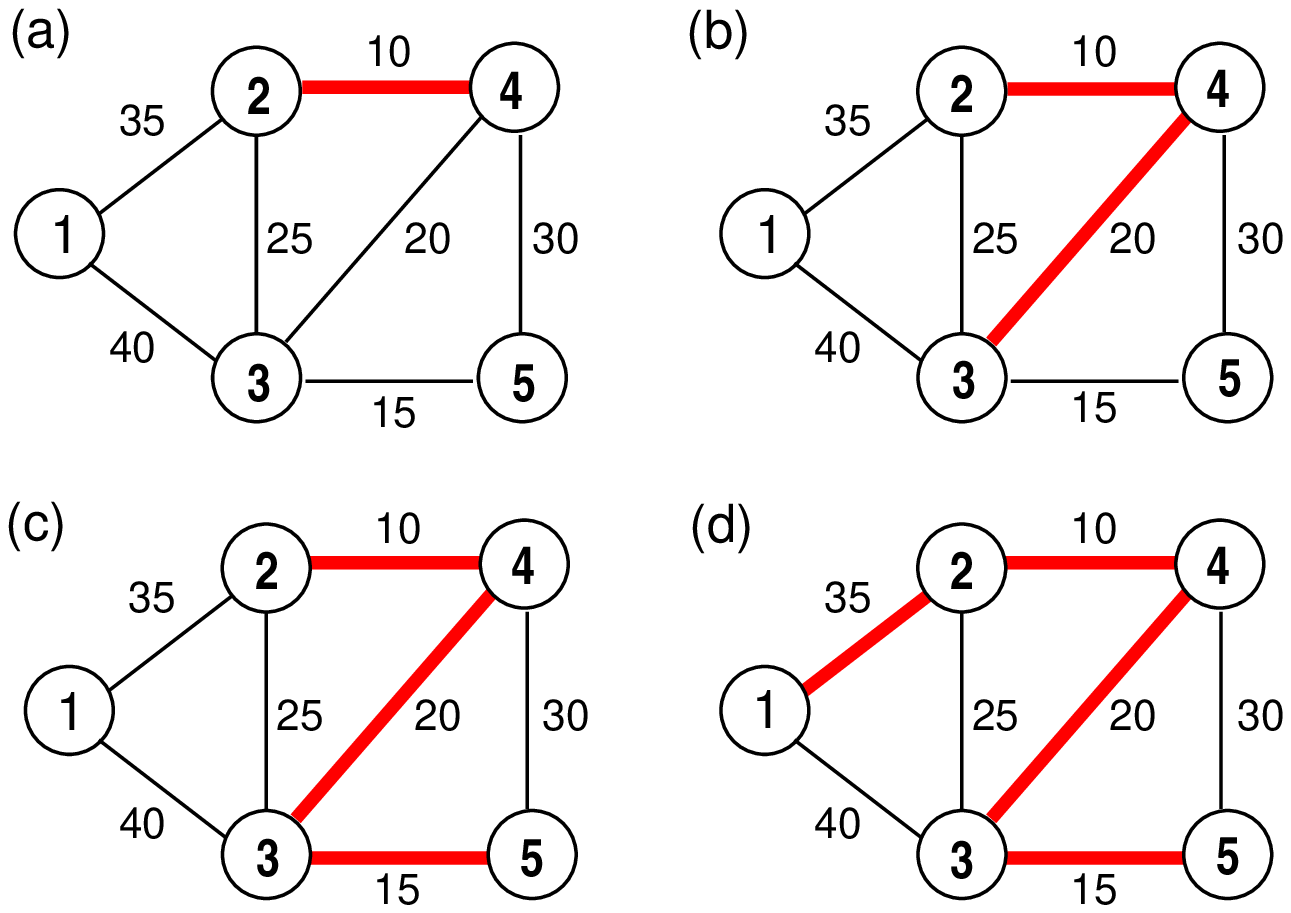}}
\end{center}
\vchcaption{
  Illustration of Prim's algorithm. The graph contains five vertices
  numbered from 1 to 5. The numbers along the edges denote the edge
  weights. The current edges contained in the spanning tree are shown in
  bold. From (a), starting with the lowest
  cost edge, to (d), successively new edges are added until a minimum
  spanning tree is reached.}
\label{fig:prim}
\end{vchfigure}
\index{Prim's algorithm|)}\index{algorithm!Prim's|)}
\index{minimum spanning tree|)}\index{spanning tree!minimum|)}
\index{graph!algorithms|)}\index{algorithm!graph|)}


\section{Random graphs}
\label{sec:3.2}
\index{random graph|(}\index{graph!random|(}

\subsection{Two ensembles}
\index{graph!ensemble|(}\index{ensemble!graph|(}

The basic idea of random-graph theory is to study \emph{ensembles}
of graphs instead of single, specific graph instances. Using
probabilistic tools, it is sometimes easier to show the existence of
graphs with specific properties, or, vice versa, to look at how a 
\emph{typical} graph \index{typical graph}\index{graph!typical}  
with given properties (e.\,g., order, size, degrees\ldots)
appears.

The simplest idea goes back to a seminal paper published by Erd\H{o}s and
R\'enyi \index{graph!Erdos-renyi@Erd\H{o}s--R\'enyi}
\index{Erdos-renyi random graph@Erd\H{o}s--R\'enyi random graph}
in 1960 \cite{ErRe}. They assumed all graphs of the same order $N$ and
size $M$ to be equiprobable. There are mainly two slightly different ensembles
used in the field:

The ensemble ${\cal G}(N,M)$ contains all graphs of $N$ vertices and
$M$ edges. The measure is flat, i.\,e., every graph has the same
probability, see, e.\,g., Fig.~\ref{fig:rg}. Note that all graphs obtained
from a given graph by permuting the vertices are different graphs, i.\,e.,
they are counted individually.

\begin{vchfigure}[htb]
\begin{center}
\vspace{0.8cm}
\scalebox{0.5}{\includegraphics{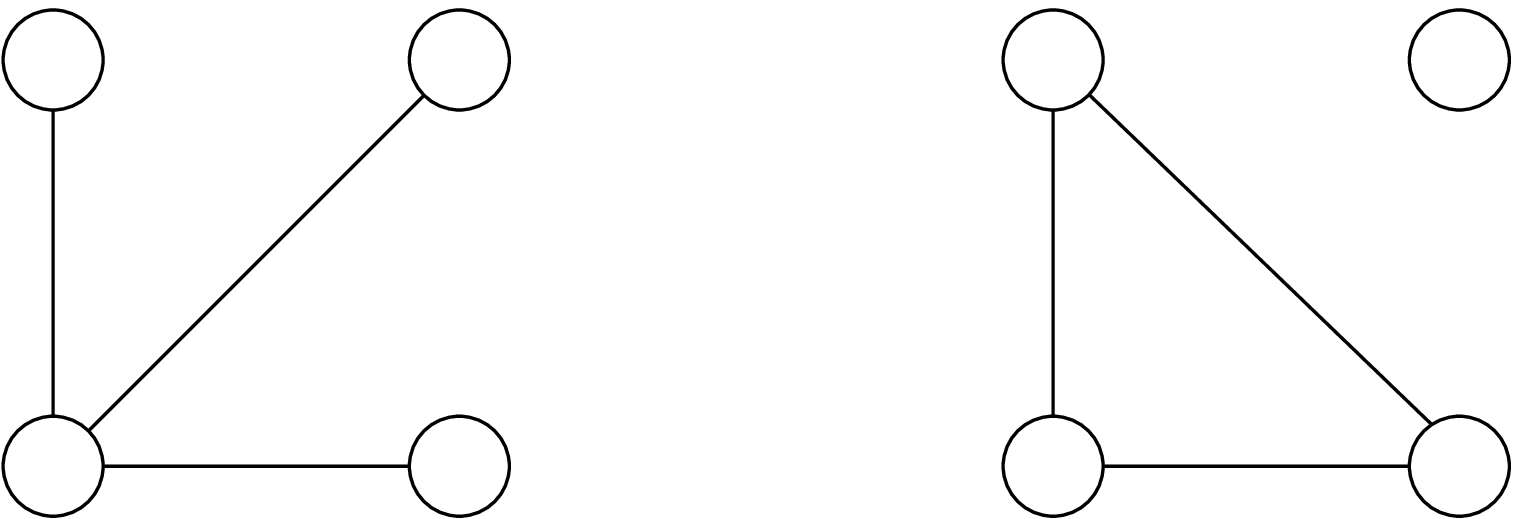}}
\end{center}
\vchcaption{These two graphs are equiprobable in ${\cal G}(4,3)$, even if
  their structure is quite different. Whereas the left graph is a
  connected tree, the right one is not connected and contains a cycle.}
\label{fig:rg}
\end{vchfigure}

The ensemble ${\cal G}(N,p)$, with $0\leq p \leq 1$, contains graphs
with $N$ vertices. For every vertex pair $i,j$ an edge $\{i,j\}$ is drawn
independently with probability $p$. For $p=0$, the graph has no edges,
for $p=1$, the graph is complete ($K_N$). On average, the number of
edges for given $p$ is
\begin{equation}
  \label{eq:edges}
  \overline M = p {N \choose 2} = p\frac{N!}{(N-2)!2!} = p\frac{N(N-1)}{2}
\end{equation}
where the over-bar denotes the average over ${\cal G}(N,p)$. This
ensemble is analytically easier to handle, so we mainly work with
${\cal G}(N,p)$.

The specific case ${\cal G}(N,1/2)$ can also be considered as 
\emph{the} random ensemble of graphs: All graphs of $N$ vertices are
equiprobable, independently of their edge numbers.

One important type of statement is that a ${\cal G}(N,p)$-graph 
fulfils \emph{almost surely} \index{almost surely} 
(or \emph{with probability one}) some condition $C$. This
expression means that, for $N\to \infty$, the probability that a graph
drawn from ${\cal G}(N,p)$ fulfils this condition $C$, converges to one.
\index{graph!ensemble|)}\index{ensemble!graph|)}

\subsection{Evolution of graphs}
\label{sec:evolution}
\index{evolution of graphs|(}\index{graph!evolution|(}

Sometimes, it is very instructive to imagine the ensemble ${\cal
G}(N,p)$ \index{graph!ensemble}\index{ensemble!graph}
via an evolutionary process of graphs of $N$ vertices which,
with  time, get more and more edges. This can be realized in
the following way. We take $V=\{1,\ldots,N\}$, and for all $i,j\in V$,
$i<j$, we independently draw a random number $x_{ij}$ being equally
distributed in $(0,1)$. Initially the graph has no edges. 
Now, we start to grow $p(t)$. Whenever it
exceeds a number $x_{ij}$, an edge between vertices $i$ and $j$ is
added. Thus, at time $t$, the graph belongs to ${\cal
G}(N,p(t))$. There are some interesting stages in this evolution:
\begin{itemize}
\item $p(t)\sim 1/N^{2}$: The first isolated edges 
\index{isolated edge}\index{edge!isolated} appear.

\item $p(t)\sim 1/N^{3/2}$: The first vertices have degree 2, i.\,e., the
  first edges have a common vertex. \index{degree}

\item $p(t)\sim 1/N^{\alpha},$ any $\alpha>1$: The graph is almost
  surely a forest. \index{forest}

\item $p(t)\sim 1/N$: The average vertex degree stays finite for $N\to
  \infty$, first cycles appear, \index{cycle}
the first macroscopic (i.\,e., of order
  ${\cal O}(N)$) subgraph \index{macroscopic subgraph}
  \index{subgraph!macroscopic} appears,  macroscopic $q$-cores 
\index{macroscopic q-core@macroscopic $q$-core}%
\index{q-core@$q$-core!macroscopic} appear.

\item $p(t)\simeq \ln(N)/N$: The graph becomes connected.
\index{connected graph}\index{graph!connected}

\item $p(t)\simeq (\ln(N)+\ln(\ln(N))/N$: The graph becomes
  Hamiltonian. \index{Hamiltonian graph}\index{graph!Hamiltonian}
\end{itemize}
For the proof see the book by Bollob\'as mentioned on page 
\pageref{page:literature}.
\index{evolution of graphs|)}\index{graph!evolution|)}

\subsection[Finite-connectivity graphs: The case $p=c/N$]{Finite-connectivity graphs: The case \for{toc}{$p=c/N$}\except{toc}{$\bm{p=c/N}$}}
\label{sec:3.2.3}
\index{finite-connectivity random graph|(}
\index{random graph!finite-connectivity|(}

The most interesting case is given by $p=c/N$, where the medium size
of the graph $\overline M = c(N-1)/2$ grows linearly with the graph
order $N$. In this section, we first discuss the fluctuations of the
total edge number $M$, and some local properties like degrees and the
existence of small cycles. In the following two sections we finally
switch to global properties. We discuss \emph{random-graph percolation}
\index{random graph!percolation}\index{percolation!random-graph}%
which describes a phase transition \index{phase transition}
from graphs having only small
connected components, even for $N\to\infty$, to graphs  also having one
giant component which unifies a finite fraction of all vertices,
i.\,e., whose order also grows linearly with the graph order $N$. The
last subsection discusses the sudden emergence of a $q$-core.

\subsubsection{The number of edges}

For a graph from ${\cal G}(N,c/N)$, every pair of vertices becomes
connected by an edge with probability $c/N$, and remains unlinked with
probability $1-c/N$. In contrast to the ${\cal G}(N,M)$-ensemble, the
total number of edges is a random variable and fluctuates from sample
to sample. Let us therefore calculate the probability $P_M$ of it having
exactly $M$ edges. \index{distribution!of edges}
\index{edge!number of $\cdot$s}\index{number of edges}%
This is given by
\begin{equation}
  \label{eq:edgenumber}
  P_M = {N(N-1)/2 \choose M} \left[ \frac cN \right]^M 
  \left[ 1- \frac cN \right]^{N(N-1)/2-M}\ .
\end{equation}
The combinatorial prefactor describes the number of possible
selections of the $M$ edges out of the $N(N-1)/2$ distinct vertex
pairs, the second factor gives the probability that they are in fact
connected by edges, whereas the last factor guarantees that there are
no further edges. In the limit of large $N$, and for $M\sim N$, where
$M\ll N(N-1)/2$, we use
\begin{eqnarray}
  \label{eq:approx}
  \left( \frac{N(N-1)}2 \right)
  \left( \frac{N(N-1)}2 -1 \right)\cdots
  \left( \frac{N(N-1)}2 - M +1 \right) 
  = &&\nonumber\\
 \left( \frac{N(N-1)}2 \right)^M 
  + {\cal O}\left(N^{2(M-1)}\right) &&
\end{eqnarray}
and $(1-c/N)^{zN}\approx \exp(-cZ)$,
hence the above expression can be asymptotically approximated by
\begin{equation}
  \label{eq:edgenumber2}
  P_M \simeq \frac{(N(N-1)/2)^M}{M!} \left[ \frac cN \right]^M 
  \exp( - c(N-1)/2)\ .
\end{equation}
Plugging in $\overline M=c(N-1)/2$, this asymptotically leads
to a \emph{Poissonian distribution} \index{Poissonian distribution|ii}
\index{distribution!Poissonian|ii}  with mean $\overline M$,
\begin{equation}
  \label{eq:edgenumber3}
  P_M \simeq \exp\left( - \overline M\right) \frac{\overline M^M}{M!} 
  \ .
\end{equation}
The fluctuations \index{fluctuation!number of edges}
\index{number of edges!fluctuations}
of the edge number $M$ can be estimated by the 
\emph{standard deviation} \index{standard deviation} of $P_M$,
\begin{equation}
  \label{eq:sdev}
  \sigma(M) = \sqrt{ \overline{(M-\overline M)^2}} = 
  \sqrt{ \overline{M^2}-\overline M^2}\ .
\end{equation}
We first calculate
\begin{eqnarray}
  \label{eq:M2}
  \overline{M^2} &=& \overline M + \overline{M(M-1)} \nonumber\\
  &=& \overline M + \sum_{M=0}^\infty M(M-1)P_M \nonumber\\
  &=& \overline M + \exp\left(-\overline M\right) \sum_{M=2}^\infty
  \frac{\overline M^M}{(M-2)!} \nonumber\\
  &=& \overline M + \overline M^2 \exp\left(-\overline M\right)
  \sum_{m=0}^\infty \frac{\overline M^m}{m!}\nonumber\\
  &=& \overline M + \overline M^2\,. 
\end{eqnarray}
In the third line, we have eliminated the cases $M=0,1$ which vanish
due to the factor $M(M-1)$, in the fourth line we have introduced
$m=M+2$ which runs from 0 to $\infty$. The sum can be performed and
gives $\exp(\overline M)$. Using Eq.~(\ref{eq:sdev}) we thus find
\begin{equation}
  \label{eq:sdev2}
  \sigma(M) = \sqrt{\overline M}\ ,
\end{equation}
which is a special case of the \emph{central limit theorem}.
\index{central limit theorem}\index{theorem!central limit}
The relative fluctuations $\sigma(M)/\overline M = 1/\sqrt{
  \overline M}$ decay to zero for $\overline M = c(N-1)/2 \to \infty$,
see Fig.~\ref{fig:edges}.
In this limit, the sample-to-sample fluctuations of the edge number
become less and less important, and the ensemble ${\cal G}(N,c/N)$ can
be identified with ${\cal G}(N,\overline M)$ for most practical 
considerations.

\begin{vchfigure}[htb]
\begin{center}
\vspace{0.8cm}
\scalebox{0.4}{\includegraphics{pic_graphs/edges.eps}}
\end{center}
\vchcaption{The distribution of edge numbers for $\overline
  M=10,100,1000$, rescaled by $\overline M$. The distributions
  obviously sharpen for increasing $\overline M$.}
\label{fig:edges}
\end{vchfigure}

\subsubsection{The degree distribution}

Let us now discuss the degrees of the graph. We are interested in the
probability $p_d$ that a randomly chosen vertex has exactly degree
$d$. The set of all $p_d$ is called the \emph{degree distribution}.
\index{degree distribution}\index{distribution!of degrees}%
It can be easily calculated:
\begin{equation}
  \label{eq:Nd}
  p_d = {N-1 \choose d} \left[ \frac cN \right]^d \left[ 1- \frac cN
  \right]^{N-d-1} \ .
\end{equation}
The meaning of the terms on the right-hand side is the following:
The factor ${N-1 \choose d}$ enumerates all possible selections for $d$
potential neighbors. Each of these vertices is connected to the
central vertex with probability $c/N$, whereas the other $N-d-1$
vertices are not allowed to be adjacent to the central one, i.\,e., they 
contribute a factor $(1-c/N)$ each. In the large-$N$ limit, where any
fixed degree $d$ is small compared with the graph order $N$, we can 
continue in an analogous way to the last subsection and find
\begin{eqnarray}
  \label{eq:pd}
  p_d &=& \lim_{N\to \infty} \frac { (N-1)(N-2)\cdots(N-d)}{N^d}
    \left[ 1- \frac cN \right]^{N-d-1} \frac {c^d}{d!}
    \nonumber\\
  &=& e^{-c} \frac {c^d}{d!}\ ,
\end{eqnarray}
i.\,e., also the degrees are distributed according to a  Poissonian
\index{Poissonian distribution} \index{distribution!Poissonian}%
distribution. It is obviously normalized, and the average degree is
\begin{eqnarray}
  \label{eq:avdegree}
  \sum_{d=0}^{\infty} d\ p_d &=& \sum_{d=1}^{\infty} e^{-c} \frac
  {c^d}{(d-1)!} \nonumber\\
  &=& c\ .
\end{eqnarray}
This was clear since the expected number of neighbors to any vertex is
$p(N-1)\to c$. Note also that the fluctuation of this value are again given by
the standard deviation $\sigma(c)=\sqrt c$, which, however, remains comparable
to $c$ if $c={\cal O}(1)$. The degree fluctuations between vertices thus also
survive in the thermodynamic limit, there is, e.\,g., a fraction $e^{-c}$ of
all vertices which is completely isolated.

If we randomly select an edge and ask for the degree of one of its
end-vertices, we obviously find a different probability
distribution $q_d$, e.\,g., degree zero cannot be reached ($q_0=0$). This
probability is obviously proportional  to $p_d$ as well as to $d$, by
normalization we thus find
\begin{equation}
  \label{eq:qd}
  q_d = \frac {d p_d}{\sum_d d p_d} =\frac {d p_d}c 
  = e^{-c} \frac{c^{d-1}}{(d-1)!}
\end{equation}
for all $d>0$. The average degree of vertices selected in
this way equals $c+1$, it includes the selected edge together with, on
average, $c$ additional \emph{excess edges}. \index{excess edge}
\index{edge!excess} The number $d-1$ of these
additional edges will be denoted as the \emph{excess degree}
\index{excess degree}\index{degree!excess}  of the
vertex under consideration.

\subsubsection{Cycles and the locally tree-like structure}

The degrees are the simplest local property. We may go slightly beyond
this and ask for the average number of small subgraphs. Let us start
with subtrees of $k$ vertices which thus have $k-1$ edges. We
concentrate on labeled subtrees \index{tree!number of $\cdot$s} 
\index{subtree!number of $\cdot$s}
(i.\,e., the order in which the vertices
appear is important) which are not necessarily induced (i.\,e., there may be
additional edges joining vertices of the tree which are not counted).
Their expected number is proportional to
\begin{eqnarray}
  \label{eq:subtree}
  N(N-1)\cdots(N-k+1) \left[ \frac cN\right]^{k-1}
  &=&   N c^{k-1} + {\cal O}(1)\ ,
\end{eqnarray}
combinatorial factors specifying $k-1$ specific edges out of the
$k(k-1)/2$ possible ones, are omitted. This number thus grows linearly with
the graph order $N$. If, on the other hand, we look to cycles of
finite length $k$, we have to find $k$ edges. The above expression
takes an additional factor $c/N$, the expected number of cycles is
thus of ${\cal O}(1)$, i.\,e., the number of triangles, squares, etc.
stays finite even if the graph becomes infinitely large! This becomes
more drastic if one looks to cliques $K_k$ with $k\geq 4$, these
become more and more likely to be completely absent if $N$ becomes
large. Thus, if we look locally to induced subgraphs of any finite
size $k$, these are almost surely trees or forests. This property is
denoted as \emph{locally tree-like}.

There are, however, loops, \index{loop}
but these are of length ${\cal O}(\ln N)$ \index{loop!length of} 
\cite{GRAPH-bollobas1985}. In the
statistical physics approach, we will see that these loops are of
fundamental importance, even if they are of diverging length.

Another side remark concerns the dimensionality of the graph, 
\index{dimension of graph}\index{graph!dimension of} i.\,e., the
possibility of ``drawing'' large graphs in a finite-dimensional space. If
you look to any $D$-dimensional lattice, the number of neighbors up to
distance $k$ grows as $k^D$. On the other hand, in a tree of fixed
average degree, this number is growing exponentially! In this sense,
random graphs have to be considered to be infinite dimensional. This
sounds strange at first, but finally explains the
analytical tractability which will be observed later in this book. 

\subsection{The phase transition: Emergence of a giant component}
\index{phase transition!percolation|(}
\index{percolation!phase transition|(}

From Sec.~\ref{sec:evolution} we know that random graphs with
$p=c/N$ are almost surely not connected. What can we say about the
components?

Having in mind the growth process described above, we may imagine that for
small $c$ there are many small components, almost all of them being trees. If
$c$ increases, we add new edges and some of the previously disconnected
components now become connected. The number of components decreases, the
number of vertices (order) of the components grows. Concentrating on the
largest component $L^{(0)}(G)$ of a graph $G$, the following
theorem was demonstrated in 1960 by Erd\H{o}s and R\'enyi \cite{ErRe}:

\textbf{Theorem:} \emph{Let $c>0$ and $G_c$ drawn from ${\cal G}(N,c/N)$.
Set $\alpha = c-1-\ln c$. \\
(i) If $c<1$ then we find almost surely 
\begin{equation*} |L^{(0)}(G_c)| = \frac 1\alpha \ln N + {\cal O}(\ln \ln N)\end{equation*}
(ii) If $c>1$ we almost surely have
\begin{equation*} |L^{(0)}(G_c)| = \gamma N + {\cal O}(N^{1/2})\end{equation*}
with $0<\gamma=\gamma(c)<1$ being the unique solution of
\begin{equation*} 1-\gamma = e^{-c \gamma}\ .\end{equation*} 
All smaller components \index{small component}\index{component!small}
are of order ${\cal O}(\ln N)$.}

This theorem makes a very powerful statement: As long as $c<1$, all
components have order up to ${\cal O}(\ln N)$. This changes markedly
if $c>1$. There appears one \emph{giant component} \index{giant component}
\index{component!giant} connecting a finite
fraction of all vertices, but all the other components are still small
compared to the giant one, they have only ${\cal O}(\ln N)$ vertices.
This means that at $c=1$ the system undergoes a \emph{phase
transition}. In contrast to physical phase transitions, it is not
induced by a change in temperature, pressure or other \emph{external}
control parameters, but by the change of the average vertex degree
$c$, i.\,e., by a \emph{structural} parameter of the graph. Due to the
obvious analogy to percolation theory \cite{St}, this phase transition
is also called \emph{random-graph percolation} 
\index{graph!percolation} in the 
literature.

This theorem is one of the most fundamental results of random-graph
theory, but we do not present a complete proof. There is, however, a
simple argument that the component structure changes at $c=1$. It is
based on the locally tree-like structure 
\index{locally tree-like structure}\index{locally tree-like component}
\index{structure!locally tree-like}\index{component!locally tree-like}
 of all components, and is
presented in the limit $N\to\infty$.

If we select any vertex, on average it will have $c$ neighbors. According to
Eq.~(\ref{eq:qd}), each of these will have $c$ additional neighbors,
the first vertex thus has, on average, $c^2$ second
neighbors. Repeating this argument, we conclude that the expected
number of $k$th neighbors equals $c^k$ (A vertex $j$ is called a $k$th
neighbor of a vertex $i$ if the minimum path in the graph $G$
connecting both contains exactly $k$ edges).

Now we can see the origin of the difference for $c<1$ and $c>1$. In
the first case, the prescribed process decreases exponentially,
and it is likely to die out after a few steps. If, in contrast, we
have $c>1$, the expected number of $k$th neighbors grows
exponentially. Still, there is a finite probability that the process
dies out after a few steps (e.\,g., $e^{-c}$ if dying in the first step,
i.\,e., if there are no neighbors at all), but there is also a finite
probability of proceeding for ever. 

\begin{vchfigure}[htb]
\begin{center}
\scalebox{0.6}{\includegraphics{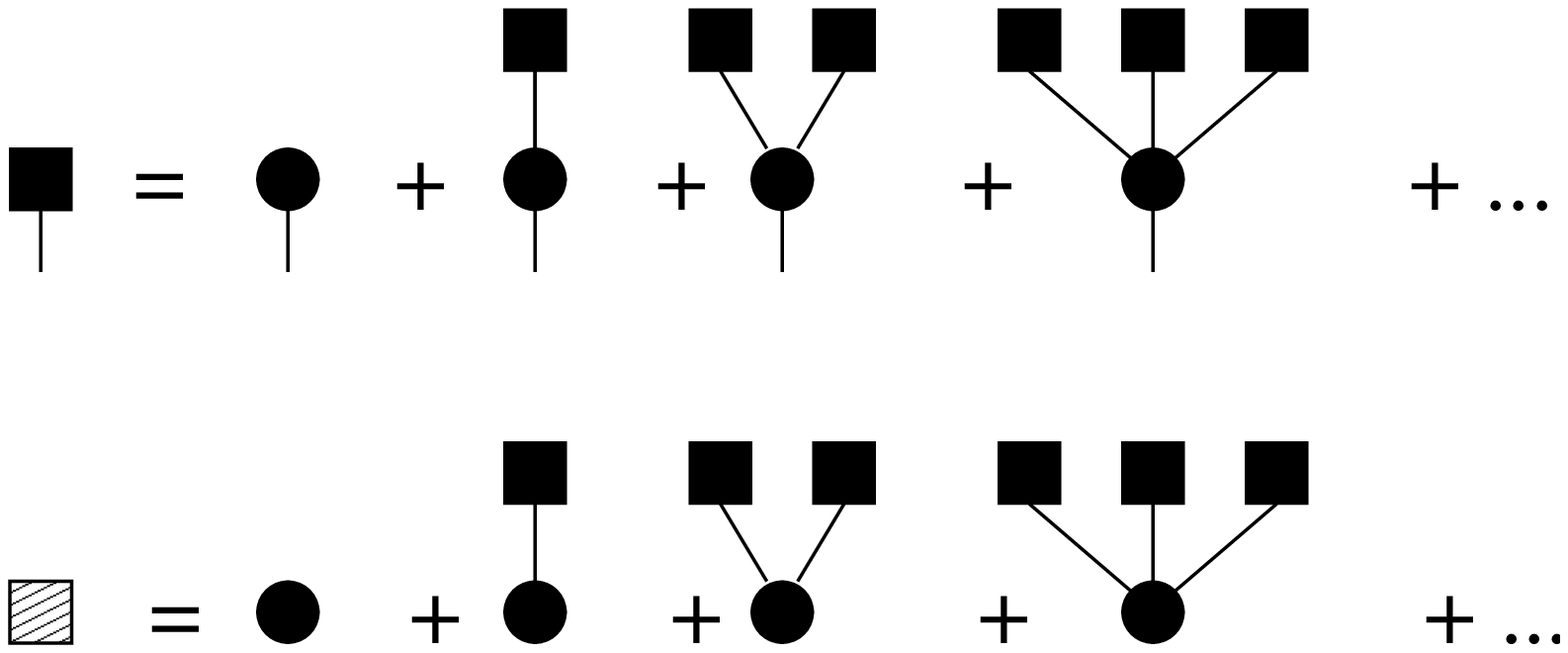}}
\end{center}
\vchcaption{Schematic representation of the iterative solution for the
probability that a vertex does not belong to the giant component.  The
first line shows the self-consistent equation for a vertex reached by
a random edge: The black square with the half-edge represents the
probability that the  vertex reached is not connected to the giant
component by one of its excess edges. This can happen because it has
no further neighbors, or it is connected to one, two, etc., vertices not
 connected with the giant component. The second line shows the
resulting equation for a randomly selected vertex, as represented by
the shaded square.}
\label{fig:iteration}
\end{vchfigure}

\index{giant component!order of|(}\index{order!of giant component|(}
This argument can be made more rigorous by considering the following
iterative construction: Fist we calculate the probability $\pi$, that
a randomly selected end-vertex of a randomly selected link is not
connected via other edges with the giant component of the graph. In
Figure~\ref{fig:iteration} this is represented by a black square
connected to a half-edge. This vertex can either have degree one,
i.\,e., it has no further incident edges which would be able to connect
it to the giant component, or it is connected to other vertices which
themselves are not connected to the giant component by their excess
edges. Having in mind that almost all small connected components are
trees, these neighbors are connected with each other only via the
selected vertex, and the probability that $d$ neighbors are not
connected to the giant component equals simply $\pi^d$. Using the
probability distribution $q_d$ Eq.~(\ref{eq:qd}) 
of degrees of vertices reached by
random edges, we thus find:
\begin{eqnarray}
  \label{eq:iter_pi}
  \pi &=& q_1 + q_2 \pi + q_3 \pi^2 + q_4 \pi^3 + \ldots \nonumber\\
  &=& \sum_{d=1}^\infty e^{-c} \frac{c^{d-1}}{(d-1)!} \ \pi^{d-1} 
=  e^{-c}\sum_{d'=0}^\infty \frac{(c\pi)^{d'}}{(d')!}
  \nonumber\\
  &=& e^{-c(1-\pi)}\ .
\end{eqnarray}
This quantity can thus be determined self-consistently for every value
of $c$, and allows us to calculate the probability $1-\gamma$ that a
randomly selected vertex does not belong to the giant
component. Applying similar arguments to before, this vertex can
either be isolated, or connected only to other vertices which, via
their excess edges, are not connected to the giant component, see the
second line of Fig.~\ref{fig:iteration}. We thus find
\begin{eqnarray}
  \label{eq:iter_gamma}
  1-\gamma &=& p_0 + p_1 \pi + p_2 \pi^2 + p_3 \pi^3 + \ldots \nonumber\\
  &=& \sum_{d=1}^\infty e^{-c} \frac{c^d}{d!} \ \pi^d 
  \nonumber\\
  &=& e^{-c(1-\pi)}\ .
\end{eqnarray}
From these equations we see first that, for our random graph,
$\pi=1-\gamma$. Plugging this into the last line, we obtain the
equation
\begin{equation}
  \label{eq:result_giant}
  1-\gamma = e^{-c\gamma}
\end{equation}
as given in the theorem.

\begin{vchfigure}[htb]
\begin{center}
\vspace{0.8cm}
\scalebox{0.4}{\includegraphics{pic_graphs/perc.eps}}
\end{center}
\vchcaption{Fraction of vertices belonging to the largest component of a
  random graph, as a function of the average degree $c$. The symbols
  are numerical data for $N=100,1000$, averaged always over 100
  randomly generated graphs. The full line gives the asymptotic
  analytical result: Below $c=1$, the largest component is
  sub-extensive, above it is extensive.}
\label{fig:perc}
\end{vchfigure}

Let us shortly discuss the shape of $\gamma(c)$. For
$c=1+\varepsilon$, with $0<\varepsilon\ll 1$, we also expect $\gamma$
to be small, and expand the above equation to second order in
$\gamma$:  
\begin{equation}
  \label{eq:critical}
  1-\gamma = 1 - (1+\varepsilon)\gamma + \frac 12 (1+\varepsilon)^2
  \gamma^2 + {\cal O}(\gamma^3) \,.
\end{equation}
The term $1-\gamma$ cancels on both sides. Dividing by $\gamma$, and
keeping only the first order in $\varepsilon$ and $\gamma$, we thus
find
\begin{equation}
  \label{eq:critical1}
  \gamma = 2 \varepsilon\ .
\end{equation}
The relative order of the giant component thus starts to grow linearly
in $c-1$ (we say the critical exponent \index{critical exponent}
for the giant component equals
one), and converges exponentially fast to 1 for larger $c$. The full
curve is given in Fig.~\ref{fig:perc}, together with numerical data
for finite random graphs.
\index{giant component!order of|)}\index{order!of giant component|)}
\index{phase transition!percolation|)}
\index{percolation!phase transition|)}

\subsection[The emergence of a giant $q$-core]{The emergence of a giant \for{toc}{$q$}\except{toc}{$\bm{q}$}-core}
\index{giant q-core@giant $q$-core|(}\index{q-core@$q$-core!giant|(}
\index{percolation!q-core@$q$-core|(}\index{q-core@$q$-core!percolation|(}

In Sec.~\ref{subsec:coloring}, 
we have introduced the $q$-core of a graph as the
maximal subgraph having minimum vertex degree of at least $q$. Quite
recently, the problem of the existence of a $q$-core in random graphs
was solved in a beautiful paper by B. Pittel, J. Spencer and
N.~C. Wormald \cite{PiSpWo} by analyzing exactly the linear-time
algorithm given there. Here we will give an easily
accessible summary of their approach, not including the technical
details of the mathematical proofs.

Let us first start with a numerical experiment: We have generated
medium-size random graphs with $N=50\,000$ vertices and various average
degrees $c$, and we have applied the $q$-core algorithm for $q=2$ and
$q=3$. The results for $q=2$ are not very astonishing: Up to $c=1$, the
2-core \index{2-core}
is very small, whereas it starts to occupy a finite fraction of
all vertices for $c>1$. This is consistent with the results on the
giant component obtained in the last section: As long as the latter
does not exist, almost all vertices are collected in finite trees, and
thus do not belong to the 2-core. A small difference in the emergence
of the giant component can, however, be seen looking in the vicinity
of $c=1$. Whereas the giant component starts to grow linearly with
the distance from the critical point, see Eq.~(\ref{eq:critical1}),
the 2-core emerges with zero slope, see Fig.~\ref{fig:core}. 
\index{phase transition!continuous}\index{continuous phase transition}%
 This means
that the critical exponents \index{critical exponent} of both
processes appear to differ from each other.

The situation changes dramatically for $q\geq 3$. \index{3-core}
At the corresponding
threshold $c(q)$, which is monotonously increasing with $q$, the
$q$-core appears \emph{discontinuously} --  
\index{phase transition!discontinuous}%
\index{discontinuous phase transition} immediately unifying a finite
fraction of all vertices. For $q=3$, the threshold is found to be
$c(3)\simeq3.35$. Slightly below this average degree, almost all
graphs have no extensive $q$-core at all. At the transition, the order
of the $3$-core jumps abruptly to about $0.27 N$, i.\,e., it contains
about 27\% of all vertices!

\begin{vchfigure}[htb]
\begin{center}
\vspace{0.8cm}
\scalebox{0.4}{\includegraphics{pic_graphs/core.eps}}
\end{center}
\vchcaption{Fraction of vertices belonging to the $q$-core for $q=2$
  (circles) and $q=3$ (diamonds). The numerical data result from a
  single graph of $N=50\,000$ vertices. The vertical line gives the
  theoretical value of the 3-core size at the threshold $c=3.35.$} 
\label{fig:core}
\end{vchfigure}

\subsubsection{Wormald's method}
\label{subsec:Wormald}
\index{Wormald's method|(}\index{method!Wormald's|(}

The method of describing this behavior was presented mathematically  by
Wormald \cite{Wo}, it is, however, well-known and applied in physics
under the name of \emph{rate equations}. \index{rate equations}
It analyzes a slightly
modified version of the $q$-core algorithm. In every algorithmic step,
only one vertex of degree smaller than $q$ is selected randomly and
removed from the graph. The quantity calculated in the analysis of
Pittel \textit{et al.} is the degree distribution $p_d$, as it changes under
the algorithmic decimation. This is obviously a perfect candidate for
determining the halting point of the algorithm. If $p_d=0$ for all
$d<q$, the remaining subgraph is the $q$-core.

An important point in the analysis, which will not be justified
mathematically here, is that the graph dynamics is \emph{self-averaging} 
\index{self-averaging} in the thermodynamic limit $N\to\infty$. After a
certain number $T=tN={\cal O}(N)$ of algorithmic steps almost all
random graphs have the same degree distribution, which is also almost
surely independent of the random order of decimated
vertices. Technically, this allows us to average over both kinds of
randomness and to determine the average, or expected action of the
algorithm.

In the last paragraph, we have already introduced a rescaling of the
number of algorithmic steps by introducing the \emph{time}
\index{time!rescaled}\index{rescaled time}
$t=T/N$. This quantity remains finite even in the thermodynamic limit,
running initially from  $t=0$ to at most $t=1$, where the full graph
would have been removed. Further on, this quantity advances in steps
$\Delta t = 1/N$, and thus becomes continuous in the large-$N$
limit. The last observation will allow us to work with differential
equations instead of discrete-time differences.

After these introductory remarks we can start the analysis by first
noting that the total vertex number evolves as $N(T)=(1-t)\cdot N$,
since we remove one vertex per step. Introducing later on the numbers
$N_d(T)$ of vertices having $d$ neighbors at time $t=T/N$, and the
corresponding degree distribution $p_d(t) = N_d(T)/N(T) $, we can
write down the \emph{expected} change of $N_d$ in the $(T+1)$st step.
It contains two different contributions:
\begin{itemize}
\item The first contribution concerns the removed vertex itself. It
  appears only in the equations for $N_d$ with $d<q$, because no
  vertex of higher current degree is ever removed.
\item The second contribution results from the neighbors of the
  removed vertex. Their degree decreases by one.
\end{itemize}
We thus find the equation \index{degree!rate equations}
\index{rate equations!of degree}%
below where we explain the meaning of the terms in detail:
\begin{equation}
  \label{eq:rate1}
  N_d(T+1)-N_d(T) = - \frac{\chi_d p_d(t)}{\overline\chi} +
  \frac{\overline{d\chi}}{\overline\chi}
  \left[ -\frac{ d\,p_d(t)}{c(t)} + \frac{ (d+1)p_{d+1}(t)}{c(t)}
    \right] 
\end{equation}
which are valid for all degrees $d$.  Here we have introduced $\chi_d$ as an
indicator for vertices of degree at most $q-1$, i.\,e., it equals one if $d<q$
and zero else. The overbar denotes as before the average over the graph,
\change{i.\,e., $\overline{\chi}=\sum_d\chi_d p_d(t)$ and
  $\overline{d\chi}=\sum_d d\chi_d p_d(t)$.  Note that these averages are now,
  even if not explicitly stated, time dependent. We keep the explicit notation
  for $c(t)\equiv \overline d =\sum_d d p_d(t)$.}
\change{All terms have the following form: they are products of how
  often the corresponding process happens on average in one step, 
and of the probability that
the corresponding process affects a vertex of degree $d$.}
The first term of the right-hand side of Eq.~(\ref{eq:rate1}) describes
the removal of the selected vertex \change{(i.\,e., this process happens exactly
once within each step)}, the indicator $\chi_d$ guarantees
that only vertices of degree $d<q$ can be removed. The second term
contains the neighbors: On average, there are
${\overline{d\chi}}/{\overline\chi}$ neighbors, each of them having
degree $d$ with probability $q_d(t)=dp_d(t)/c(t)$. A loss term stems
from vertices having degree $d$ before removing one of their incident
edges, a gain term from those having degree $d+1$.

In the thermodynamic limit, the difference on the left-hand side of
Eq.~(\ref{eq:rate1}) becomes a derivative:
\begin{eqnarray}
  \label{eq:diff}
  \label{eq:rate}
  N_d(T+1)-N_d(T) &=& \frac{(1-t+\Delta t)p_d(t+\Delta t) - 
  (1-t)p_d(t)}{\Delta t} \nonumber\\ 
  &=& \frac d{dt} \left\{(1-t)p_d(t) \right\}\ .
\end{eqnarray}

The last two equations result in a closed set of equations for the
degree distribution,
\begin{equation}
  \label{eq:gr_rate}
  \frac d{dt} \left\{(1-t)p_d(t) \right\} = 
  - \frac{\chi_d p_d(t)}{\overline\chi} +
  \frac{\overline{d\chi}}{\overline\chi}
  \left[ -\frac{ d\,p_d(t)}{c(t)} + \frac{ (d+1)p_{d+1}(t)}{c(t)}\ ,
    \right] 
\end{equation}
which will be solved in the following. First we derive some global
equations for averages over $p_d(t)$. The simplest one can be obtained
by summing Eq.~(\ref{eq:gr_rate}) over all $d$, which gives the trivial
consistency relation $\frac d{dt} (1-t) = -1$. A much more interesting
equation results by first multiplying Eq.~(\ref{eq:gr_rate}) by $d$,
and then summing over all $d$. Introducing the short-hand notation
$m(t)=(1-t) c(t)$, we find
\begin{eqnarray}
  \label{eq:gr_m}
  \dot m(t) &=& \frac d{dt} \left\{ (1-t) 
  \sum_d d p_d(t)\right\} \nonumber\\
  &=& \frac{\overline{d\chi}}{\overline\chi}
  \ \left[ -1 - \frac{ \overline{d^2}}{c(t)}
    +  \frac{ \overline{d(d-1)}}{c(t)} \right] \nonumber\\
  &=& - 2 \frac{\overline{d\chi}}{\overline\chi} \ .
\end{eqnarray}
\change{The initial condition is hereby $m(t=0)=c$}.

The main problem in solving Eqs~(\ref{eq:rate}) is, however, that they form
an infinite set of non-linear differential equations. A direct solution is
therefore far from obvious. Our salvation lies in the fact that the algorithm
never directly touches a vertex of degree $d\geq q$. These vertices change
their properties only via the random removal of one of their incident edges.
Therefore they maintain their random connection structure, and their
Poissonian shape, only the effective connectivity changes.  This can be
verified using the ansatz, with $\beta(t)$ being the effective connectivity
($\beta(0) = c$):
\begin{equation}
  \label{eq:poisson_ansatz}
  (1-t) p_d(t) = \frac{N_d(T)}{N} 
\stackrel{!}{=} e^{-\beta(t)} \frac{\beta(t)^d}{d!}\ ,\qquad
  \forall\  d\geq q\,.
\end{equation}
This ansatz introduces a radical simplification.
An infinite number of probabilities is parametrized by one single
parameter $\beta(t)$. The validity of this ansatz can be justified by
plugging it into Eq.~(\ref{eq:rate}) for arbitrary $d$. We obtain
for the left-hand side
\begin{equation}
  \label{eq:lhs}
  {\rm l.h.s.} = \dot \beta(t) \left[
  e^{-\beta(t)} \frac{\beta(t)^{d-1}}{(d-1)!} 
  -e^{-\beta(t)} \frac{\beta(t)^d}{d!} 
  \right]
\end{equation}
and for the right-hand side
\begin{equation}
  \label{eq:rhs}
  {\rm r.h.s.} = \frac{\overline{d\chi}}{\overline\chi}
  \left[ -e^{-\beta(t)}\frac{\beta(t)^d}{(d-1)! (1-t) c(t)} 
    +e^{-\beta(t)}\frac{\beta(t)^{d+1}}{d! (1-t) c(t)}
    \right] \ .
\end{equation}
They have to be equal for all $d\geq q$, and we can compare the
components of the monomial of the same order of both sides:
\begin{equation}
  \label{eq:beta}
  \dot \beta(t) = - \frac{\beta(t)}{m(t)}
  \frac{\overline{d\chi}}{\overline\chi}\ .
\end{equation}
The equations thus become closed under ansatz (\ref{eq:poisson_ansatz}), and
instead of having an infinite number of equations for the $p_d(t)$, $d=q,q+1,\ldots,$ just one equation for $\beta(t)$ remains. Interestingly,
the $q$-dependence in this equation is dropped.

If we compare Eqs~(\ref{eq:gr_m}) and (\ref{eq:beta}), we find
\begin{equation}
  2\frac{\dot\beta(t)}{\beta(t)} = \frac{\dot m(t)}{ m(t)}
\end{equation}
which, using the initial conditions, is solved by
\begin{equation}
  \label{eq:mb}
  m(t) = \frac{\beta(t)^2}c\ .
\end{equation}
The function $m(t)$ can thus be replaced in Eq.~(\ref{eq:beta}), and
we get
\begin{equation}
  \label{eq:beta2}
  \dot \beta(t) = - \frac{c}{\beta(t)}
  \frac{\overline{d\chi}}{\overline\chi}\ .
\end{equation}

As a further step, we still have to remove $\overline{\chi}$ and
$\overline{d\chi}$ from the last equation. This can be obtained by
using normalization, when applying Eq.~(\ref{eq:poisson_ansatz})
\begin{eqnarray}
  \label{eq:norm_chi}
  1 &=& \sum_{d=0}^\infty p_d(t) \nonumber\\ 
  &=& \sum_{d=0}^\infty \chi_d p_d(t) + 
  \sum_{d=q}^\infty \frac 1{1-t} e^{-\beta(t)} \frac{\beta(t)^d}{d!}
  \nonumber\\
  &=& \overline \chi + \frac 1{1-t} F_q(\beta(t))\,.
\end{eqnarray}
In the same way, we obtain for the average degree $c(t)$:
\begin{equation}
  \label{eq:F_q}
  F_q(\beta) := 1-\sum_{d=0}^{q-1} e^{-\beta} \frac{\beta^d}{d!}\ ,
\end{equation}
and
\begin{eqnarray}
  \label{eq:deg_chi}
  c(t) &=& \sum_{d=0}^\infty d p_d(t) \nonumber\\ 
  &=& \sum_{d=0}^\infty d \chi_d p_d(t) + 
  \sum_{d=q}^\infty \frac d{1-t} e^{-\beta(t)} \frac{\beta(t)^d}{d!}
  \nonumber\\
  &=& \overline{d\chi} + \frac {\beta(t)}{1-t} F_{q-1}(\beta(t))\ .
\end{eqnarray}
Inserting this into Eq.~(\ref{eq:beta2}), and using Eq.~(\ref{eq:mb})
to eliminate $c(t)=m(t)/(1-t)$, the equation for $\beta(t)$ finally
becomes:
\begin{equation}
  \label{eq:beta3}
  \dot\beta(t) = -\frac{ \beta(t) -c F_{q-1}(\beta(t)) }{
     1-t - F_{q}(\beta(t)) }\ .
\end{equation}
We have thus succeeded in reducing the infinite set (\ref{eq:gr_rate})
into the single Eq.~(\ref{eq:beta3})! Still, this equation is
non-linear and thus not easily solvable for arbitrary $q$. The
important information about the $q$-core can, however, be obtained
even without solving this equation for all times, but by determining
only the halting point of the algorithm -- which is the $q$-core of
the input graph.

Assume that the graph has a giant $q$-core. It contains $N(t_f) =
(1-t_f) N$ vertices, where $0<t_f<1$ is the halting time of the
algorithm. At this point, all remaining vertices have degree $d\geq
q$, we thus have $\overline\chi=\overline{d\chi}=0$ and $\beta_f =
\beta(t_f)>0$. According to (\ref{eq:norm_chi},\ref{eq:deg_chi}) we
therefore have
\begin{eqnarray}
  \label{eq:halt}
  1-t_f &=& F_q(\beta_f) \nonumber\\
  \frac{\beta_f}c &=& F_{q-1}(\beta_f) \ .
\end{eqnarray}
The second equation fixes $\beta_f$, whereas $t_f$ itself and thus the
order of the giant $q$-core follow directly from the first
equation. If there are two or more solutions for $\beta_f$ one
has to choose the largest one which is smaller than $c$. The initial
condition for $\beta(t)$ is $\beta(0)=c$, and Eq.~(\ref{eq:beta3})
results in a negative slope, i.\,e., in a decrease of $\beta(t)$ with time.

\subsubsection{The case $\bm{q=2}$}
\index{2-core|(}

Let us first consider the case of $q=2$ colors only. From numerical
experiments we have seen that the 2-core of a random graph seems to
set in continuously at average degree $c=1$. Can we confirm this,
starting with Eqs~(\ref{eq:halt})?

The self-consistent equation for $\beta_f=\beta(t_f)$ becomes
\begin{equation}
  \label{eq:2core_beta}
  \frac{\beta_f}c  = 1 - e^{-\beta_f}\ ,
\end{equation}
as represented graphically in Fig.~\ref{fig:2core_res}. The right-hand
side starts in $\beta=0$ with slope one, and asymptotically tends to
one. The left-hand side is a straight line of slope $1/c$. For $c<1$
the only solution of Eq.~(\ref{eq:2core_beta}) is thus given by
$\beta_f=0$, which according to the first of Eqs~(\ref{eq:halt})
corresponds to $1-t_f=0$, and the 2-core has vanishing size. At $c>1$,
the right-hand side has a smaller slope in the origin, but still
diverges for large $\beta$. The two curves therefore have a second
crossing point $\beta_f>0$ which, according to the discussion at the
end of the last section, has to be selected as the relevant
solution. A non-zero $\beta_f$, however, implies a non-zero $1-t_f$,
and the 2-core unifies a finite fraction of all vertices of the random
graph.

\begin{vchfigure}[htb]
\begin{center}
\vspace{0.8cm}
\scalebox{0.4}{\includegraphics{pic_graphs/2core.eps}}
\end{center}
\vchcaption{Graphical solution of Eq.~(\ref{eq:2core_beta}): The full
  line shows $F_1(\beta)$, the straight lines have slopes $1/c_i$ with
  $c_1<c_2=1<c_3$. The solution is given by the largest crossing point
  between the curves. For $c\leq 1$ the only crossing appears in
  $\beta=0$.  For $c>1$, a second, positive solution exists. It is
  marked by the arrow. }
\label{fig:2core_res}
\end{vchfigure}

Note that the non-zero solution of Eq.~(\ref{eq:2core_beta}) sets in
continuously at $c(2)=1$. If we fix a slightly larger
$c=1+\varepsilon$ with $0<\varepsilon\ll 1$, we find
\begin{equation}
  \label{eq:2core_crit}
  \frac{\beta_f}{1+\varepsilon}  = 1 - e^{-\beta_f}\ .
\end{equation}
Expanding this equation in $\beta_f$ and $\varepsilon$ we get
\begin{equation}
  \label{eq:2core_crit2}
  \beta_f(1-\varepsilon+{\cal O}(\varepsilon^2) ) = 
  \beta_f\left(1-\frac {\beta_f} 2 +{\cal O}(\beta_f^2)\right)
\end{equation}
and thus $\varepsilon=\beta_f/2$. For the halting time, and thus for the
size of the 2-core, we have
\begin{eqnarray}
  \label{eq:2core_tf}
  1-t_f 
  &=& 1- e^{-\beta_f} (1+\beta_f) \nonumber\\ 
  &=& \frac{\beta_f^2}2+ {\cal O}(\beta_f^3)
  \nonumber\\
  &=& 2 \varepsilon^2 + {\cal O}(\varepsilon^3)\ .
\end{eqnarray}
The critical exponent \index{critical exponent} 
for 2-core percolation is thus $2$, which is in
perfect agreement with the numerical findings presented at the
beginning of this section.
\index{2-core|)}

\subsubsection{The case $q=3$}
\index{3-core|(}

For $q=3$, the self-consistency equation for $\beta_f=\beta(t_f)$
becomes
\begin{equation}
  \label{eq:3core_beta}
  \frac{\beta_f}c  = 1 - e^{-\beta_f} (1+\beta_f)\ ,
\end{equation}
see Fig.~\ref{fig:3core_res} for the graphical solution. The important
difference to the 2-core is that, for $q=3$, the right-hand side
starts with slope zero. A non-zero crossing point developing
continuously out of the solution $\beta=0$ is therefore impossible.
In fact, up to $c(3)\simeq 3.35$, no further crossing point exists,
and the existence of a giant component of the random graph is
obviously not sufficient for the existence of a giant 3-core. At
$c(3)\simeq 3.35$, however, the two curves touch each other at one
point $\beta_f>0$, as marked by the full-line arrow in
Fig.~\ref{fig:3core_res}, and \emph{two} new solutions appear. The
larger one is, as already discussed, the correct one.

\begin{vchfigure}[htb]
\begin{center}
\vspace{0.8cm}
\scalebox{0.4}{\includegraphics{pic_graphs/3core.eps}}
\end{center}
\vchcaption{Graphical solution of Eq.~(\ref{eq:3core_beta}): The full
  line shows $F_2(\beta)$, the straight lines have slopes $1/c_i$ with
  $c_1<c_2=3.35<c_3$. The solution is given by the largest crossing
  point between the curves. For $c\leq 3.35$ the only crossing appears
  in $\beta=0$.  For $c>3.35$, two other, positive solutions
  exist. The correct one is marked by the arrows.}
\label{fig:3core_res}
\end{vchfigure}

The discontinuous emergence of a new solution in
Eq.~(\ref{eq:3core_beta}) results also in a discontinuous emergence of
the 3-core. As observed in the experiments, the latter jumps at
$c(3)\simeq 3.35$ from zero size to about 27\% of all vertices.  This
becomes even more pronounced if we look to larger $q$, e.\,g., the 4-core
\index{4-core} appears at $c(4)\simeq 5.14$ and 
includes $0.43 N$ vertices, and the
5-core \index{5-core} 
emerges only at $c(5)\simeq 6.81$ and includes even $0.55 N$
vertices.
\index{3-core|)}
\index{Wormald's method|)}\index{method!Wormald's|)}
\index{giant q-core@giant $q$-core|)}\index{q-core@$q$-core!giant|)}
\index{percolation!q-core@$q$-core|)}\index{q-core@$q$-core!percolation|)}

\subsubsection{Results for random-graph coloring}
\index{random graph!coloring of $\cdot$|(}
\index{coloring!of random graphs|(}

As an immediate consequence we see that almost all graphs with
$c<c(q)$ are colorable with $q$ colors, or, formulated differently,
that the chromatic number for a graph with $c<c(q)$ is almost surely
bounded from above by $q$. It may, of course, be smaller. We know that
the existence of a $q$-core is not sufficient to make a graph
uncolorable with $q$ colors, see Fig.~\ref{fig:col}. As noted in the
proof of the \change{theorem on page \pageref{page:coloring}, 
the reversion of the $q$-core
algorithm can be used to extend the coloring of the $q$-core to the
full graph} in linear time. So
we see that, even if the problem is hard in the worst case, almost all
graphs of $c<c(q)$ can be efficiently colored with only $q$ colors.

The analysis given by Pittel, Spencer, and Wormald is very similar to an
analysis of a simple linear time algorithm that we give in
Chap.~8.  In general, the probabilistic analysis of
algorithms is able to provide useful bounds on properties of almost all random
graphs.
\index{random graph!coloring of $\cdot$|)}
\index{coloring!of random graphs|)}
\index{finite-connectivity random graph|)}
\index{random graph!finite-connectivity|)}
\index{random graph|)}\index{graph!random|)}
\index{graph|)}


\end{document}